\documentclass[%
superscriptaddress,
 aps,
 amsmath,amssymb,
 noeprint,
 pra,
 twocolumn
]{revtex4-2}

\usepackage{bm}
\usepackage[utf8]{inputenc}
\usepackage{amsthm}
\usepackage{graphicx}
\usepackage{amsfonts}
\usepackage{dsfont}
\usepackage{ulem}
\usepackage{float}
\usepackage{makecell}
\usepackage{braket}
\usepackage{fancyvrb}
\usepackage{siunitx}
\usepackage{blkarray}
\usepackage{hyperref}
\usepackage{soul}
\usepackage{xcolor}
\newcommand{\bs}[1]{\boldsymbol{#1}}

\newcommand{\HarvardPhysics}{Department of Physics, Harvard University, Cambridge, MA 02138, USA}
\newcommand{\IBM}{IBM Quantum, IBM T.J. Watson Research Center, Yorktown Heights, NY 10598 USA}
\newcommand{\NVIDIA}{NVIDIA, Santa Clara, California 95051, USA}
\newcommand{\BU}{Department of Physics, University of Colorado Boulder/JILA, Boulder, CO 80309, USA}
\newcommand{\fml}{f_{\text{ML}}}
\newcommand{\heff}{H^{\text{input}}_{\text{eff}}}
\newcommand{\new}[1]{{\color{black} #1}}
\newcommand{\framing}[1]{{\color{black} #1}}

\makeatletter
\def\maketitle{
\@author@finish
\title@column\titleblock@produce
\suppressfloats[t]}
\makeatother

\begin{document}

\title{\framing{Expressive Quantum Perceptrons for Quantum Neuromorphic Computing}}
\date{\today}

\author{Rodrigo Araiza Bravo}
\email{oaraizabravo@g.harvard.edu}
\affiliation{\HarvardPhysics}
\author{Taylor L. Patti}
\affiliation{\NVIDIA}
\author{Khadijeh Najafi}
\affiliation{\IBM}
\affiliation{\HarvardPhysics}
\author{Xun Gao}
\affiliation{\HarvardPhysics}
\affiliation{\BU}
\author{Susanne F. Yelin}
\affiliation{\HarvardPhysics}%

\begin{abstract}
Quantum neuromorphic computing (QNC) is a sub-field of quantum machine learning (QML) that capitalizes on inherent system dynamics. As a result, QNC can run on contemporary, noisy quantum hardware and is poised to realize challenging algorithms in the near term. One key issue in QNC is the characterization of the requisite dynamics for ensuring expressive quantum neuromorphic computation. We address this issue by proposing a building block for QNC architectures, what we call quantum perceptrons (QPs). Our proposed QPs compute based on the analog dynamics of interacting qubits with tunable coupling constants. We show that QPs are, with restricted resources, a quantum equivalent to the classical perceptron, a simple mathematical model for a neuron that is the building block of various machine learning architectures. \framing{Moreover, we show that QPs are theoretically capable of producing any unitary operation.} Thus, QPs are computationally more expressive than their classical counterparts. As a result, QNC architectures built our of QPs are, theoretically, universal. We introduce a technique for mitigating barren plateaus in QPs called entanglement thinning. We demonstrate QPs' effectiveness by applying them to numerous QML problems, including calculating the inner products between quantum states, energy measurements, and time-reversal. Finally, we discuss potential implementations of QPs and how they can be used to build more complex QNC architectures.
\end{abstract}

\maketitle

\section{Introduction}\label{Sec:Intro}
Classical machine learning often employs simple mathematical models for neuron dynamics as computational building blocks~\cite{hornik1989multilayer}. Recent work in quantum machine learning (QML) algorithm design illuminates the similarities between the dynamics of neurons in the brain and those of noisy qubits~\cite{martinez2020information, PRXQuantum.3.030325}. These studies, named quantum neuromorphic computing (QNC)~\cite{markovic2020quantum, mujal2021opportunities}, promise to deliver noise-resilient, scalable algorithms for QML implementations in the near-term. An outstanding question in QNC is finding simple, scalable models for neuron dynamics which serve as the building blocks of quantum neuromorphic computing. \par 
Indeed, QNC using superconducting qubits has already produced quantum versions of memory units called quantum memory resistors (i.e., quantum memristors)~\cite{pfeiffer2016quantum}. These quantum memristors realize quantum versions of realistic neuronal models such as the Hodgkin-Huxley model~\cite{gonzalez2019quantized, gonzalez2020quantized}. However, the Hodgkin-Huxley model is too complex to serve as a scalable building block. A more straightforward approach aims to quantize a mathematical model inspired by single neuron dynamics such as the perceptron~\cite{freund1998large}. Currently, there are several proposed quantum perceptrons, but they rely on computational architectures that are inaccessible in the near-term~\cite{kapoor2016quantum, kerstin2019efficient}, slow adiabatic computation~\cite{torrontegui2019unitary,ban2021speeding,pechal2022direct}, or multiqubit interactions~\cite{tacchino2019artificial,mangini2020quantum,PhysRevX.12.021037}. Proposing an experimentally tractable quantum perceptron is recognized as an important open question in QML~\cite{schuld2022quantum}. \par 
The present manuscript addresses this gap by proposing a version of a quantum perceptron (QP) based on the natural unitary dynamics of qubit ensembles equipped with time-dependent single-qubit control fields. These dynamics are sufficient to recover the classical perceptron and enable expressive quantum computation. On the other hand, a single classical perceptron cannot approximate generic functions. Classical perceptrons are universal approximators only when many perceptrons are stacked together~\cite{hornik1989multilayer}. Thus, as we propose them, a QP's computational power outshines that of the classical perceptron. \par

The structure of this manuscript is as follows. In Sec.~\ref{Sec:QP}, we define QPs via an interaction Hamiltonian and a set of time-dependent single-qubit control fields. In Sec.~\ref{Sec:Recovery}, we show that QPs, under certain limiting conditions, approximate classical perceptrons. Concretely, QPs approximate standard activation functions in ML in a noise-resilient way. Therefore, architectures built out of QPs promise to be noise-resilient. In Sec.~\ref{Sec:Expressive}, we show that the effective Hamiltonian evolution of a QP can, in theory, approximate any unitary. Thus, QPs are computationally advantageous over their classical counterpart. Sec.~\ref{Sec:Learning} describes how to use QPs for learning using a hybrid quantum-classical optimization procedure. We show that in QPs barren plateaus can be mitigated via entanglement regularization, and to this end, we introduce a technique called entanglement thinning, which eases the training of QPs. In Sec.~\ref{Sec:Applications}, we apply QPs to three important problems in quantum computing: measuring quantum state inner products, entanglement detection via energy measurement, and quantum metrology via time-reversal operations. The strong performance of QPs on these tasks establishes that their dynamics enable learning. Sec.~\ref{Sec:experimental} proposes possible hardware implementations of QPs and ways in which perceptrons can be stacked to form more intricate QNC architectures. We conclude an discuss future directions in Sec.~\ref{Sec:Conclusion}. Code to reproduce our results can be found in Ref.~\cite{QuantumPerceptrons}.

\section{Quantum Perceptrons}\label{Sec:QP}
\begin{figure*}[ht!]
    \centering
    \includegraphics[scale=0.4]{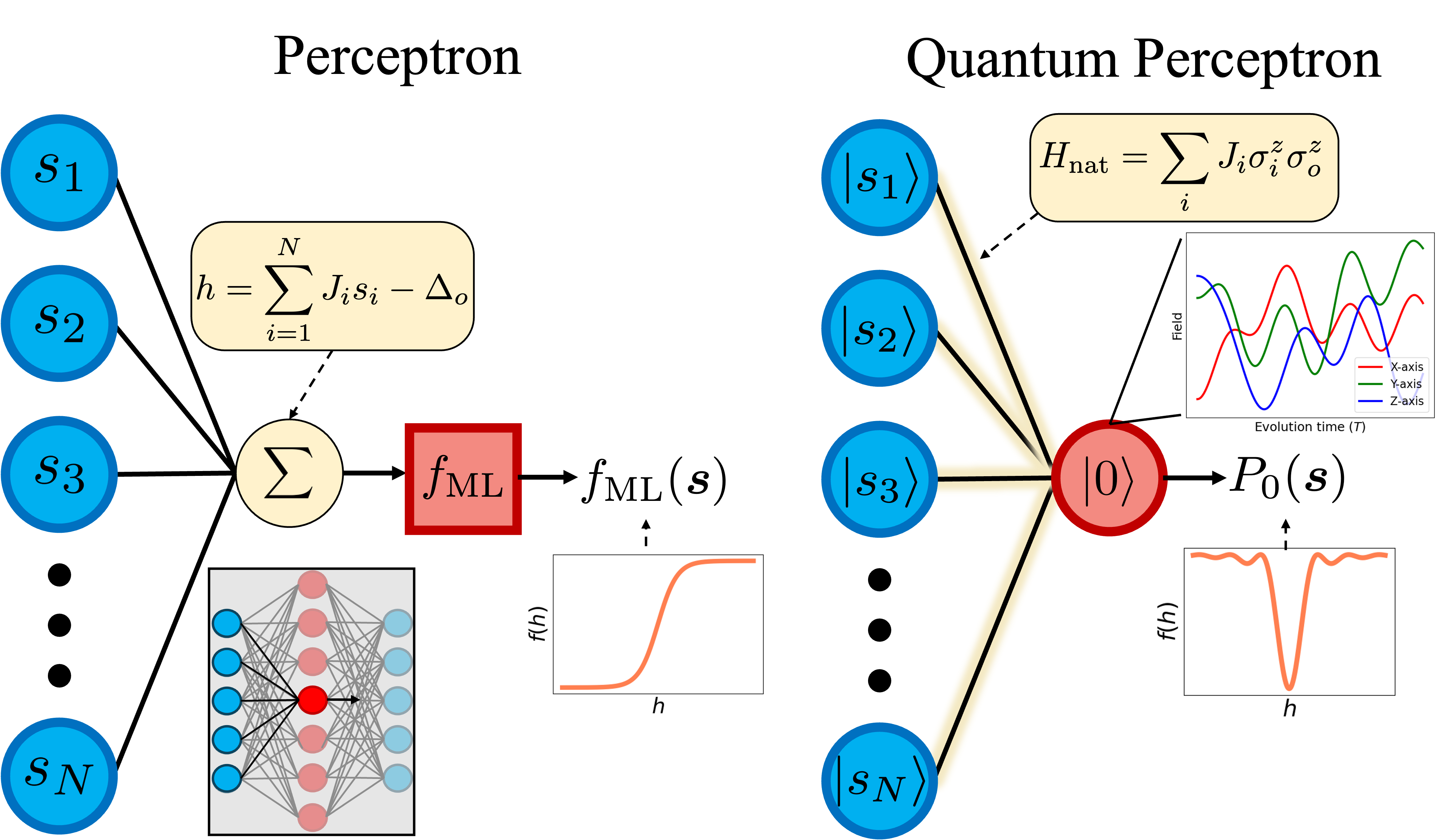}
    \caption{Schematic of perceptrons. The classical perceptron (left) operates on the input $N$-dimensional vector $\bs{s}$ to produce an output $y(\bs{s})=\fml(\sum_{i=1}^NJ_is_i-\Delta_o)$ where $J_i$ are weights and $\Delta_o$ a bias. The nonlinear function $\fml$ is called the activation function. Here we exemplify $\fml$ as the sigmoid function $\fml(h) = 1/(1+e^{-h})$. The weights and biases are chosen such that $\fml(\bs{s})$ approximates a target function $\tilde{y}(\bs{s})$, but a single perceptron cannot approximate all functions. However, as illustrated in the bottom left, neural networks composed of many perceptrons can approximate any function as long as the intermediate layer is sufficiently large. The proposed quantum perceptron (right) consists of $N$ input qubits and one output qubit. The qubits evolve under the Hamiltonian in Eq.~(\ref{eq:Hp}), which ensures that the probability $P_0(\bs{s})$ of the output qubit being in the state $\ket{0}_o$ is a nonlinear function of the input. We take $P_0(\bs{s})$ as the quantum perceptron's output. The inset on the bottom right shows the example of the activation function of a quantum perceptron as defined in Eq.~(\ref{eq:NativeF}). As specified by Eq.~(\ref{eq:Hp}), each qubit experiences time-dependent fields along the $x,y$ and $z$ directions as exemplified in the inset on the top right.}
    \label{fig:Schematic}
\end{figure*}
\subsection{Review of classical perceptrons}\label{Sec:CPs}
In machine learning, the perceptron (also called the McCulloch-Pitts neuron) is a function that decides if an input belongs to a specific class~\cite{freund1998large}. Suppose we have an input represented by a vector $\bs{s} = (s_1,...,s_N)$ where $s_i=\pm 1$. We define the perceptron as the mathematical operation 
\begin{equation}
    \bs{s}\rightarrow y(\bs{s}) = f\left(\sum_{i=1}^N J_is_i-\Delta_o\right)
\end{equation}
where $J_i$ are weights for the different features $s_i$, $\Delta_o$ a bias, and $f$ is a nonlinear function. The output is the single scalar $y(\bs{s})$. Typically, $f$ is chosen to be the sigmoid function $f(x) = 1/(1+e^{-x})$ or the rectilinear (ReLu) function where $f(x)=0$ if $x<0$ and otherwise $f(x)=x$. Fig.~\ref{fig:Schematic} shows a schematic of the perceptron (left).\par
In general, we seek weights and bias such that the perceptron approximates the desired operation $\tilde{y}(\bs{s})$ for $N_s$ samples in a training set $\{\bs{s}_i\}_{i=1}^{N_s}$. To do this, one computes the square-loss 
\begin{equation}
    \mathcal{L}(\bs{\theta}) = \frac{1}{N_s}\sum_{i=1}^{N_s}(\tilde{y}(\bs{s}_i)-y(\bs{s}_i))^2,
\end{equation}
where $\bs{\theta}$ is a vector made out of $J_i$ and $\Delta_o$. Then, we update the weights and bias via gradient descent 
\begin{equation}
    \bs{\theta}_{i} \rightarrow \bs{\theta}_{i} - \eta \frac{\partial \mathcal{L}}{\partial \bs{\theta}_{i}}
\end{equation}
where $\eta$ is the learning rate. The training set is separable if a hyperplane separates positive ($\tilde{y}(\bs{s})>0$) and negative ($\tilde{y}(\bs{s})<0$)  inputs. For such a set, there exist weights $J_i$ and bias $\Delta_o$ such that the perceptron's output converges to the correct classification, i.e., $y(\bs{s}) = \tilde{y}(\bs{s})$~\cite{novikoff1963convergence}. However, if the training set is not separable in $N$ dimensions, such convergence is not guaranteed. \par 
One of the primary goals of machine learning is producing mathematical operations that can approximate any $\tilde{y}(\bs{s})$ to arbitrary precision. If an operation can approximate any function, it is known as a universal approximator. While a single perceptron is not a universal approximator, a neural network consisting of two layers is universal as long as the first layer contains a large number of perceptrons~\cite{hornik1989multilayer}. Fig.~\ref{fig:Schematic} shows a schematic of this multilayer-perceptron (i.e., a neural network) at the bottom left.  

\subsection{Quantization of the perceptron}
We define a quantum perceptron (QP) as a set of $N+1$ qubits with $N$ inputs nodes labeled by $i=1,..., N$ and a single output node labeled by $o$. Each qubit inhabits the Hilbert space $\mathcal{H}_n$ spanned by the basis vectors $\{\ket{0}_n, \ket{1}_n\}$ such that the entire system is in the product Hilbert space $\mathcal{H} = \bigotimes_{n=1}^{N+1}\mathcal{H}_n$. For our definition, it is important to define $\sigma^x_n = \ket{0}\bra{1}_n+\ket{1}\bra{0}_n$ as the Pauli-X operator on qubit $n$, $\sigma^y_n = i\ket{0}\bra{1}_n-i\ket{1}\bra{0}_n$ as the Pauli-Y operator, and $\sigma^z_n= \ket{0}\bra{0}_n-\ket{1}\bra{1}_n$ as the Pauli-Z operator.\par 
\new{The system evolves under the time-dependent Hamiltonian depending on a set of variational parameters $\bs{\theta}$ consisting of couplings $J_i$ and field-strengths $\Delta_{nk}^{\alpha}$
\begin{align}\label{eq:Hp}
    H_{P}(t;\bs{\theta}) &= H_{\text{nat}}(\bs{J})+\sum_{n,\alpha}f_{n}^\alpha(t;\bs{\Delta})\sigma_n^\alpha,\\
    H_{\text{nat}}(\bs{J}) &= \sum_{i=1}^{N}J_{i}\sigma_i^z\sigma_o^z,\notag\\
    f_n^\alpha(t;\boldsymbol{\Delta}) &= \sum_{k=1}^L\Delta_{nk}^{\alpha}g_k(t).\notag
\end{align}
Here $H_{\text{nat}}$ is a native Hamiltonian enclosing the interactions $\boldsymbol{J}$ between qubits in the system; in this case input-output interactions only with strength $\{J_i\}$. The functions $f_i^\alpha$ represent a time-dependent field on qubit $n$ along the $\alpha\in\{x,y,z\}$ axis. These fields are linear combinations of $L$ simple  functions $g_k(t)$ which can be a set of Gaussians center at $t_k=Tk/L$ with a width $\sigma$ (i.e., $g_k(t) = (2\pi\sigma^2)^{-1/2}\exp\{-(t-t_k)^2/2\sigma^2\}$), Fourier terms with frequency $2\pi k/L$ (i.e., $g_k(t) = \cos(\frac{2\pi}{L}kt)$), or other options in from the literature~\cite{leng2022differentiable}. \par 
Lastly, we assume that the system evolves for a total time $T$ following Schrodinger's equation so that the system evolves by the unitary 
\begin{equation}
    U(T;\bs{\theta}) = \mathcal{T}e^{-i\int_0^TH_P(t;\bs{\theta})dt}
\end{equation}
where $\mathcal{T}$ is the time-ordering operator.\par
Fig.~\ref{fig:Schematic} contrast the classical and quantum perceptrons.}

\subsection{Recovering the classical perceptron}\label{Sec:Recovery}

\begin{figure}[b]
\includegraphics[width=0.95\linewidth]{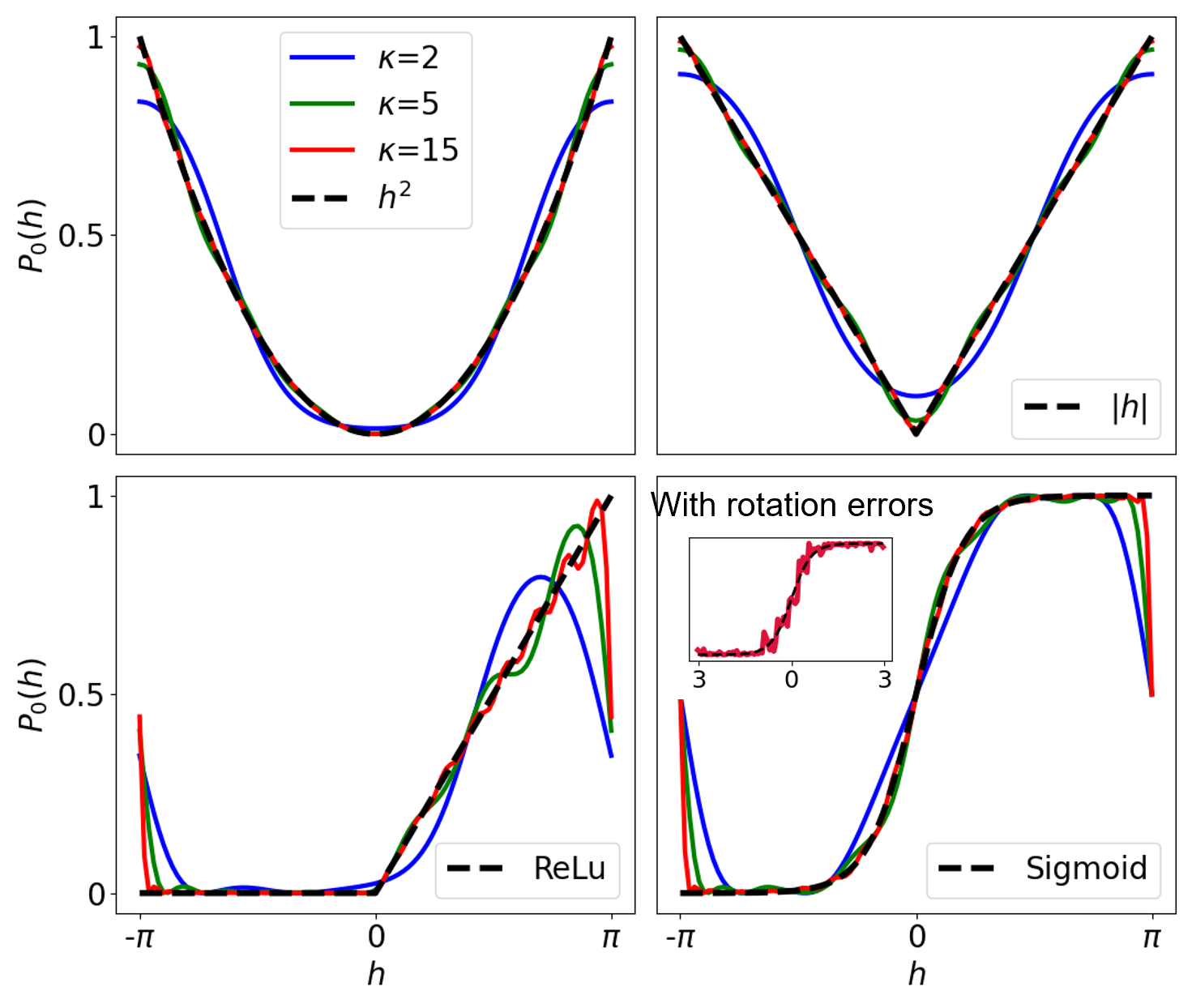}
\caption{Nonlinear functions approximated by a QP with single-qubit rotations on the output. For these approximations, we used the first 2 (blue), 5 (green), and 15 (red) Fourier coefficients of $f_{\text{ML}}$. We plot the target functions for guidance (black dotted curves). The inset shows the approximated function on the interval $[-3,3]$ with errors in both the values of the Fourier coefficients (x-rotations) and the value of the field $h$ (z-rotations) for $f_{\text{ML}}(h) = 1/(1+e^{-h})$. In Appx.~\ref{A:Pulsing}, we show that QPs are robust to noise in the field $h$. This result is due to the unitaries $W_k$ and $V_k$ in Eq.~(\ref{eq:Uf_pulsed}), where $h$ appears in conjugated operators.}
\label{fig:approximations}
\end{figure}

Let us start with a simple example of the evolutions attainable by Eq.~(\ref{eq:Hp}). This example shows that we can indeed recover the classical perceptron. Let the input qubits experience no control fields, and the output qubit have time-dependent fields only along the $x$ and $z$ axis. Under these conditions, the input qubits initialized in a state $\ket{\Psi_0}=\sum_{\bs{s}}\psi_{\bs{s}}\ket{\bs{s}}$, where $\ket{\bs{s}}$ is an eigenstate for each Pauli-Z on the input, evolves to 
\begin{equation}\label{eq:PEvolved}
    \sum_{\bs{s}}\psi_{\bs{s}}\ket{\bs{s}} \left[\cos f_{\text{QML}}(\bs{s};\bs{\theta})\ket{0}_o+i\sin f_{\text{QML}}(\bs{s};\bs{\theta})\ket{1}_o\right]
\end{equation}
where $f_{\text{QML}}(\bs{s};\bs{\theta})$ is a nonlinear function of both the string $\bs{s}$ and the parameters $\theta$ (see Eq.~(\ref{eq:NativeF}) in Appx.~\ref{A:Pulsing} for the description of $f_{\text{QML}}$). This nonlinear function is native to the QP and may not be the usual activation function $\fml$ for machine learning. Fig.~\ref{fig:Schematic} exemplifies this function. We now focus on implementing the activation functions familiar to the machine learning community. The sigmoidal function $f_{\text{ML}}(\bs{s};\bs{\theta}) = 1/(1+e^{-\sum_i J_is_i-\Delta_0})$ is a popular example~\cite{nwankpa2018activation}. \par 
A good place to start is by observing that Eq.~(\ref{eq:PEvolved}) effectively acts on the output qubit as the unitary operation $U_{f_{\text{QML}}}=e^{-if_{\text{QML}}(\bs{s};\bs{\theta})\sigma^x_o}$. That is, $U_{f_{\text{QML}}}$ evolves the output qubit as $\ket{0}_o\rightarrow \cos{f_{\text{QML}}}\ket{0}_o-i\sin{f_{\text{QML}}}\ket{1}_o$. But $f_{\text{QML}}\neq \fml$, and so the question is, could we approximate $\fml$ by operations like $U_{f_{\text{QML}}}$?\par 
To do this, we assume that the activation function we desire has a Fourier series valid on an interval $h\in [-\pi, \pi]$ where $h=\sum_iJ_is_i-\Delta^z_o$. That is, we assume that 
\begin{equation}
    \fml(h) = \frac{a_0}{2}+\sum_{k=1}^{\infty}\left(a_k\cos(kh)+b_k\sin(kh)\right)
\end{equation}
with Fourier coefficients $a_o, a_k$, and $b_k$. In Appx.~\ref{A:Pulsing}, we show that 
\begin{equation}
    U_{\fml} = e^{-ia_0\sigma^x_N}\Pi_k W_k(h)V_k(h)\label{eq:Uf_pulsed}
\end{equation}
approximates $\fml$ if
\begin{align}
    W_k(h) &= \exp\left\{-i\frac{a_k}{2}\left(e^{-i\frac{kh}{2}\sigma_o^z}\sigma_o^x e^{i\frac{kh}{2}\sigma_o^z}+\text{h.c}\right)\right\} \notag\\
    V_k(h) &= \exp\left\{i\frac{b_k}{2}\left(e^{-i(\frac{kh}{2}+\frac{\pi}{4})\sigma_o^z}\sigma_o^x e^{i(\frac{kh}{2}+\frac{\pi}{4})\sigma_o^z}+\text{h.c}\right)\right\} \notag.
\end{align}
These operators are readily realizable through the QP as follows. For example, for $W_k$, we let the output evolve under $\sigma_o^x$ with a large field amplitude $\Delta^x_o\gg J_i$ such that we can neglect the interaction. This rapid pulse should happen for a time $\tau$ such that $\Delta^x_o\tau$ is proportional to $a_k$ or $b_k$. We then let the qubits interact via $\sum_i J_i\sigma_i^z\sigma_o^z-\Delta^z_o\sigma^z_o$ for specified amounts of time, namely $\tau = k/2$. A similar process works to produce $V_k$. These two steps make the result of Eq.~(\ref{eq:Uf_pulsed}) achievable via the Hamiltonian in Eq.~(\ref{eq:Hp}) in a data-independent fashion (i.e., independet of $\bs{s}$). \par 
This calculation illustrates that if we take Eq.~(\ref{eq:Hp}) with its time-dependent, single-qubit control fields we can generate any bounded and piece-wise continuous activation function. We note that, recently, Ref.~\cite{yu2022power} introduced a similar approach to approximating a single-variable function with applications to quantum neural networks. \par 
In practice, we approximate the gate $U_{f_{\text{ML}}}$ by a finite number of Fourier terms $\kappa$. Fig.~\ref{fig:approximations} shows the results of approximating the activation functions $x^2, |x|$, sigmoid, and ReLu using $\kappa=2,5,15$ Fourier terms. The largest discrepancies with the activation functions are confined to the edges of the interval on which the Fourier series is defined. In practice, there might be errors in each evolution's duration, generating additional discrepancies from the desired activation function. The inset in Fig.~\ref{fig:approximations} shows the activation function when zero-mean Gaussian noise with standard deviation $\sigma = 0.01$ is added to $a_k, b_k,$ and $h$. We note that the duration of the pulses of the field $\Delta^x_o$ correspond to the Fourier coefficients $a_k$ and $b_k$. On the other hand, the duration of the pulses of the field $h$ corresponds to the frequencies of the Fourier series. In Appx.~\ref{A:Pulsing}, we show that QPs are robust to noise in the field $h$. This result is due to the unitaries $W_k$ and $V_k$, where $h$ appears in conjugated operators. However, the perceptron is more sensitive to noise in the field $\Delta^x_o$ since this corresponds to changing the Fourier coefficients, and thus the function $f_{\text{ML}}$ is less well approximated.\par 
Ref.~\cite{yu2022power} suggests approximating multi-variable functions by a multi-qubit system connected by CNOT gates. Our result in Eq.~(\ref{eq:Uf_pulsed}) makes this suggestion rigorous. From Eq.~(\ref{eq:Uf_pulsed}), it follows that a QP can implement any activation function with a Fourier series since we can approximate  $\fml(\bs{s}) = \fml\left(\sum_{i}J_is_i-\Delta_o\right)$. Additionally, a number $N_{out}$ of QPs can simulate multi-variable functions $f(\bs{s}) = (f_1(\bs{s}),\hdots,f_{N_{out}}(\bs{s}))$. To do this, note that output neuron $o\in \{1,\hdots,N_{out}\}$ can be chosen to experience couplings and z-field $\{J_{oi}, \Delta^z_o\}$ and single qubit rotations in Eq.~(\ref{eq:Uf_pulsed}) approximating $f_o\left(\sum_{i}J_{oi}s_i-\Delta^z_o\right)$. Notice that $H_P$ creates entanglement between the input and the output. Indeed, the CNOT gate is a special case of $H_P$ with additional single-qubit fields since $H_P$ can produce a control-Z gate (see Appx.~\ref{eq:CZinputoutput} for details). Thus, a series of perceptrons can approximate any multi-variable function as long as the function's components have a Fourier series. 

\subsection{Expressivity of Quantum Perceptrons}\label{Sec:Expressive}

\begin{figure}[b]
\includegraphics[width=0.95\linewidth]{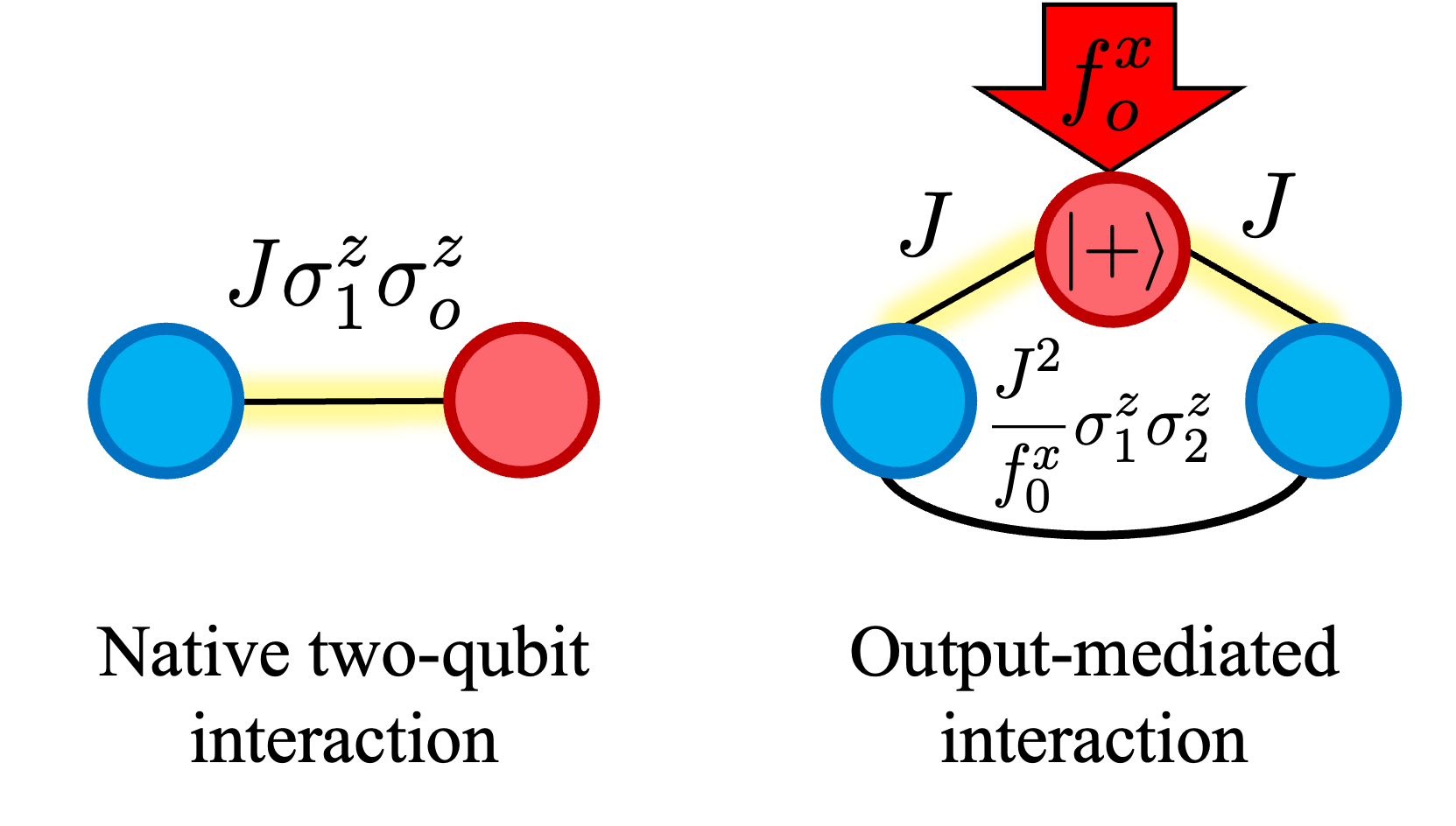}
\caption{Quantum perceptrons can realize any quantum computation. Single-qubit operations are available due to the time-dependent fields. Two qubit operations between any input and the output qubit are readily available due to the native Hamiltonian (right). Inputs can interact with each other in second-order perturbation theory where the output mediates the interaction (right).}
\label{fig:Universality}
\end{figure}
\framing{
Sec.~\ref{Sec:Recovery} illustrates how one may use a QP with time-dependent, single-qubit control fields on the output qubit to approximate single-variable functions. Similarly, a series of QPs can approximate many multi-variable functions. In other words, a QP, as described in Eq.~(\ref{eq:Hp}) with the added restrictions of only nonzero fields on the output qubit, exhibits at least the same computational complexity as a classical perceptron. In this section, we show that by fully using Eq.~(\ref{eq:Hp}) QPs can, in practice, achieves any quantum computation. That is, a QP's computational complexity exceeds that of its classical counterpart. The motivation of this section is to show QP's advantage over classical perceptrons. This does not mean QP's are an advantageous alternative, from a computational complexity and scaling perspective, to quantum circuits. Nonetheless, QP's are experimentally amenable as discussed in Sec.~\ref{Sec:experimental}.} \par 
To show that a QP can achieve any quantum computation, it suffices to establish that a restricted version of the fields in Eq.~(\ref{eq:Hp}) achieves a universal gate set. That is, QPs can do (i) any single-qubit rotation, (ii) decouple qubits in case we do not want to operate on them, and (iii) perform two-qubit gates among any pair of qubits.\par
First single-qubit rotations are straight forward. The second term in Eq.~(\ref{eq:Hp}) contains terms of the form $\bs{f}_n\cdot \bs{\sigma}_n$ where $\bs{f}_n=(f_n^x,f_n^y,f_n^z)$ and $\bs{\sigma}_n=(\sigma_n^x,\sigma_n^y,\sigma_n^z)$. If applied for a time $\tau$, these Hamiltonian would rotate qubit $n$ around the unit vector $\hat{\bs{f}}_n$ by an amount $|\bs{f}_n|\tau$. \par 
Suppose we want to rotate qubit $n$ in the QP by an amount $\theta$ around the the unit vector $\hat{\bs{f}}_n$. We can do this by chirping the time-dependent fields on that qubit such that $|\bs{f}_n|\gg J_{i}$ for a short time $\tau$ so that $|\bs{f}_n|\tau=\theta$. \par 
Second, to decouple qubits from the perceptron, we can do this by intermittently rotating them around the $x$-axis by an angle $\pi/2$. This can be done, similarly to the previous paragraph, by choosing $\bs{f}_n = f_n^x\hat{x}$ such that $f^x_n\gg J_i$ for a time $\tau = \pi/2f^x_n$. This intermittent rotation, zeros out its interaction with the output qubit (for details, see Eq.~\ref{eq:DoNothing} in Appx.~\ref{A:Universality}).\par 
Lastly, two-qubit gates among any one input (say $1$) and the output qubit is straightforward given the $J_1\sigma^z_1\sigma^z_o$ Hamiltonian term, as illustrated by the left panel in Fig.~\ref{fig:Universality}. Choosing $f^z_o=-\pi/4$, $J_1 = -\pi/2$ with $f^x_o=0$ results in the evolution 
\begin{align}\label{eq:CZinputoutput}
    \exp\left\{-i\frac{\pi}{4}\left(\sigma_i^z+\sigma_o^z-\sigma_1^z\sigma_o^z\right)\right\} &= e^{-i\pi/4}e^{i\pi\ket{1}\bra{1}_1\otimes\ket{1}\bra{1}_o} \notag\\
    &=e^{-i\pi/4}CZ_{1o}
\end{align}
where we have used the observation $2\ket{1}\bra{1}_i = \mathds{1}_i-\sigma_i^z$~\cite{gao2017quantum}. Thus, up to an arbitrary phase, a QP entangles the input and the output via a CZ gate. Moreover, the output qubit can mediate an interaction between any two inputs (say 1, and 2) which can entangle them, as illustrated by the right panel of Fig.~\ref{fig:Universality}. Input-input interactions are achieved by perturbation theory such that the inputs interact via an effective Hamiltonian 
\begin{equation}\label{eq:TwoQubitHamiltonian}
    H_{\text{nat}}^{\text{eff}}=\frac{J_1J_2}{f^x_o}\sigma^z_1\sigma^z_2
\end{equation}
See Eq.~\ref{eq:HPPerturbative} in Appx.~\ref{A:Universality}) for details. The inputs undergo a CZ-gate by evolving them under this Hamiltonian repeatedly by a total time of $\tau$ such that $\frac{J_1J_2}{f^x_o}\tau=\pi/4$, and adding single-qubit fields on the input qubits. Appx.~\ref{A:Universality}) presents a rigorous derivation of all three requirements above. \par 
This section shows that Eq.~(\ref{eq:Hp}) can, in theory, create any quantum computation. Thus, QPs are theoretically more computationally expressive than its classical counterpart. However, this manuscript isn't concern with using QPs for universal quantum computation like a quantum circuit. Instead, this proof is presented to show the theoretical advantage of QPs over classical perceptrons. Whether QPs can lead to practical advantage over classical computation or over other quantum computing paradigms is ongoing research. Research in optimal quantum control, for example, argues that variational systems like the QP may be readily, and efficiently used for producing complex unitary time evolution \cite{ho2006effective}, and calls for deepening our scholarship in this regard \cite{magann2021pulses}. \par
In what follows, we explain how to use QPs for learning, how to improve trainability, and we highlight QPs’ usefulness in quantum information processing applications.

\section{Learning with quantum perceptrons}\label{Sec:Learning}

\begin{figure*}[ht!]
\begin{center}
\includegraphics[width=\linewidth]{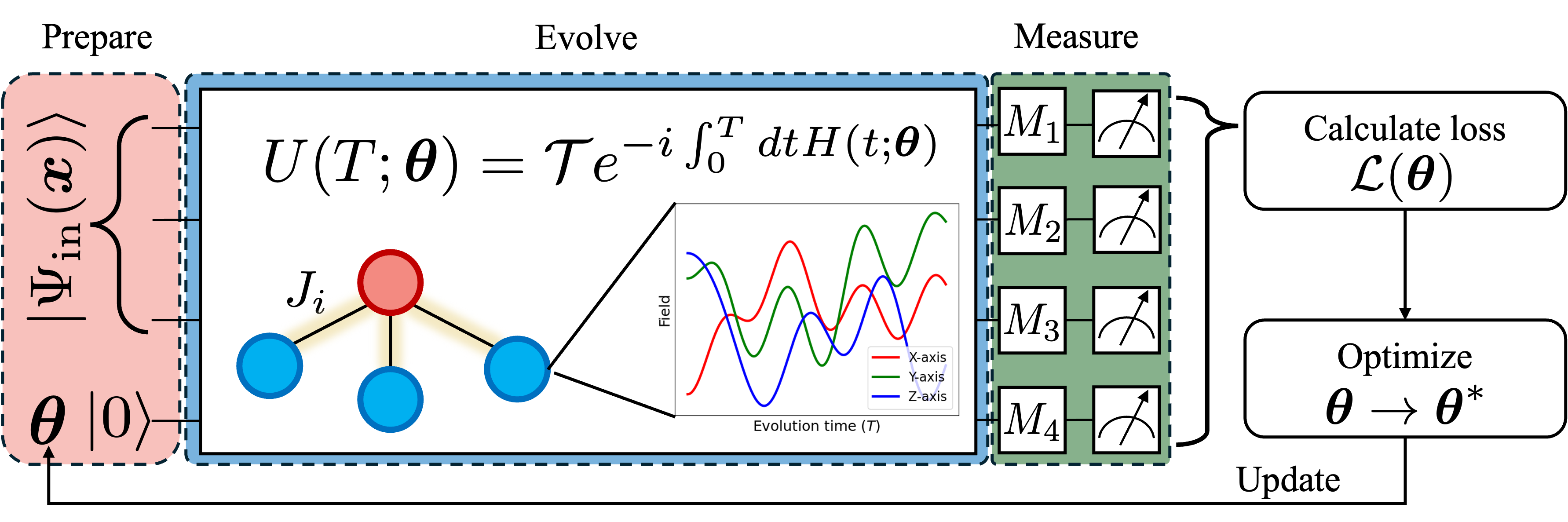}

\caption{\framing{Learning with quantum perceptrons (QP). On the first stage (left, red), we prepare an input $\bs{x}$ into the $N$ qubit state $|\Psi_{\text{in}}(\bs{x})\rangle$, and the output qubit is in state $|0\rangle$. In the second stage (center, blue), the input-output system evolves under the Hamiltonian prescribed by Eq.~(\ref{eq:Hp}) for a total time $T$. The variational parameters are $\bs{\theta}$ consisting of coupling constants $\bs{J}$ and field strengths $\bs{\Delta}$. The fields are sums of $L$ simple time-dependent functions. After the evolution, the system is measured (right, green). The measurements are used to calculate a loss function $\mathcal{L}(\bs{\theta})$. A classical optimizer such as gradient descent is then used to optimize the parameters. }}
\end{center}
\label{fig:Learning}
\end{figure*}

Thus far, we have been using QPs to realize known functions or quantum circuits. We now explore using the Hamiltonian Eq.~(\ref{eq:Hp}) as a machine-learning algorithm in a hybrid quantum-classical fashion. \par 
Recall that the unitary $U(T;\bs{\theta})$ produced by the time-dependent evolution depends on the variational parameters $\bs{J}\in \mathbb{R}^{N}$ and $\bs{\Delta}\in \mathbb{R}^{3\times (N+1)\times L}$. 
Let us now use a QP for learning. Learning with QPs consists of initializing these parameters and the system, evolving the system under $U(T;\bs{\theta})$, measuring a set of observables $\{\mathcal{O}_i\}$ and using these measurements to calculate a loss function $\mathcal{L}(\bs{\theta})$. Subsequently, a classical optimizer changes the QP's parameters so that the loss is improved. The system is re-initialized and the new parameters are then used to evolve the system. This process is repeated until the loss is optimized. See Fig.~\ref{fig:Learning} for a diagrammatic representation of the learning procedure. \par 
As an example, let us focus in the simple case where we measure the expectation value of an observable $\hat{\mathcal{O}}$, use the mean-square loss, and optimize via gradient descent. We initialize the inputs and output qubits in the state $|0\rangle^{\otimes N+1}$. The task is to minimize the loss 
\begin{equation}\label{eq:LossSingle}
    \mathcal{L}(\bs{\theta}) = \frac{1}{2}\left( \langle 0|U^\dagger(T;\bs{\theta})\hat{\mathcal{O}}U^\dagger(T;\bs{\theta})|0\rangle-\mathcal{O}_0\right)^2 = \frac{1}{2}\varepsilon(\bs{\theta})^2
\end{equation}
where $\mathcal{O}_0$ is the target value and $\varepsilon$ is the residual error. The parameters of the QP are then optimized using gradient descent 
\begin{equation}
    \bs{\theta}(t+1)=\bs{\theta}(t)-\eta\frac{\partial \mathcal{L}}{\partial\bs{\theta}(t)}
\end{equation}
where $\eta$ is the learning rate, and $t$ denotes the $t^{th}$ training epoch. \par 
Similarly, if instead of looking to obtain a single output value $\mathcal{O}_0$, we are given a data set $\mathcal{D}$ with data points $\bs{x}$ and labels $\tilde{y}(\bs{x})$, the data can be loaded into the input qubits via a feature map $\bs{x}\rightarrow |\Psi_{\text{in}}(\bs{x})\rangle|0\rangle$, where the last ket represents the output qubit. Then, the QP outputs 
\begin{equation}
 y_{\bs{\theta}}(\bs{x}) = \langle 0|\langle \Psi_{\text{in}}(\bs{x})|U^\dagger(\bs{\theta})\mathcal{O}U^\dagger(\bs{\theta})|\Psi_{\text{in}}(\bs{x})\rangle|0\rangle
\end{equation}
and the task becomes minimizing the loss
\begin{equation}\label{eq:LossMultiple}
    \mathcal{L}(\theta) = \frac{1}{2}\sum_{\bs{x}\in \mathcal{D}}\left(\tilde{y}(\bs{x})-y_{\bs{\theta}}(\bs{x})\right)^2 = \frac{1}{2}\sum_{\bs{x}\in \mathcal{D}}\varepsilon_{\bs{x}}(\bs{\theta})^2
\end{equation}
where $\varepsilon_{\bs{x}}$ is the residual error for data point $\bs{x}$.

\subsection{Mitigating Barren Plateaus in Quantum Perceptrons}\label{sec:BPs}
\begin{figure*}[ht!]
\begin{center}
\includegraphics[width=\linewidth]{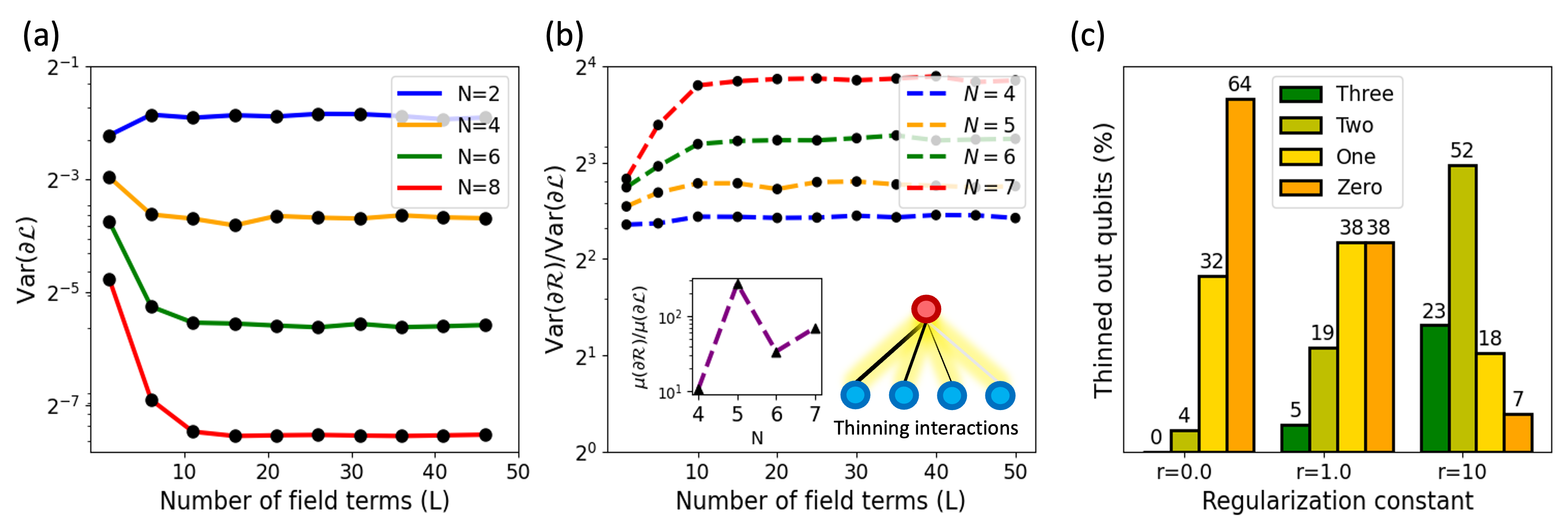}
\caption{Mitigating barren plateaus and easing the trainability of quantum perceptrons (QPs). \textbf{(a)} Shows the variance of the loss function's gradients when randomly initialized as a function of the number of field parameters $(L)$. We observe that QPs exhibit barren plateaus where the mean of $\partial \mathcal{L}$ flattens out to zero and its variance is exponentially small in the number of qubits. \textbf{(b)} Shows the effect on of entanglement regularization on the QP's barren plateaus. The variance of the gradients in the regularized case exponentially improves those found in (a). The left inset also shows that the regularized average gradients improve by one to two orders of magnitude. The right inset is a schematic representation of our proposed regularization called \textit{entanglement thinning} (ET) where interaction strengths are mitigated or reinforced (see Eq.~(\ref{eq:thinning})). \textbf{(c)} Shows the number of qubits thinned away by minimizing Eq.~(\ref{eq:CHSH}) with ET for a variety of regularizing constants $r$. For this problem, only one qubit should interact with the input. We see that $r=10$ leads to a QP with effectively one or two qubits as desired.}
\label{fig:Trainability}
\end{center}
\end{figure*}
In this section, we study a common challenge of training QPs, barren plateaus, and introduce a method to address them. As an example, we explore the trainability of a toy-problem where the loss function corresponds to forming a singlet between one input qubit and the output qubit. Mathematically, this corresponds to Eq.~(\ref{eq:LossSingle}) with the operator
\begin{equation}\label{eq:CHSH}
    \hat{\mathcal{O}}= \sigma_1^x\sigma_o^x+ \sigma_1^z\sigma_o^z,\qquad \mathcal{O}_0=-2. 
\end{equation}
In this section, we will write the loss function resulting from substituting Eq.~(\ref{eq:CHSH}) into Eq.~(\ref{eq:LossSingle}) as $\mathcal{L}(\boldsymbol{J})$. Our notation highlighting the $\boldsymbol{J}$-dependence is motivated by the fact that barren plateaus arise from the entanglement generated by the QPs evolution, as explained below. \par 
Barren plateaus (BPs) are a common issue of variational quantum algorithms in QML where randomly initialized circuits features flat training landscapes~\cite{mcclean2018barren}. Mathematically, a BP means
\begin{align}\label{eq:BPs}
\text{Avg}_{\boldsymbol{\theta}}\left(\partial_{\boldsymbol{\theta}}\mathcal{L}\right)=0,\qquad \text{Var}_{\boldsymbol{\theta}}\left(\partial_{\boldsymbol{\theta}} \mathcal{L}\right)\sim 2^{-2N}
\end{align}
where the average is taken over the randomly initialized parameters $\boldsymbol{\theta}$. BPs emerge due to over-parametrization~\cite{mcclean2018barren}, poor choices of loss functions~\cite{cerezo2021cost}, and noisy hardware~\cite{wang2021noise}. A plethora of mitigating techniques exists such as symmetry-constrained ansatze~\cite{wang2022symmetric} and entanglement regularization~\cite{patti2021entanglement}. \par 
Fig.~\ref{fig:Trainability}(a) shows the variance of the gradient $\partial \mathcal{L}(\boldsymbol{J})$ as a function of $N$ and number of field parameters ($L$). That is, the task in Eq.~(\ref{eq:CHSH}) exhibits a barren plateau. We argue that this BP results from the large entanglement build-up due to the perceptron's star-like architecture. In Appx.~\ref{A:ED} we present with numerical arguments for why entanglement is the source of the BP. Thus, entanglement mitigation promises to lift the BP.
\par 
Although we tried two entanglement mitigation techniques, here we report on the most successful one and include the other in Appdx.~\ref{A:ED}. We call this technique \textit{entanglement thinning} (ET) which aims to reduce interactions between the input and output qubits. ET has its roots in sparsifying approaches in classical machine learning such as Ridge regression~\cite{tibshirani1996regression}, and pruning~\cite{liu2015sparse}. In essence, ET amounts to regularizing the loss function via a scale-free term $H_r(\boldsymbol{J})$ (i.e., $H_r(a\boldsymbol{J})=H_r(\boldsymbol{J}$). We follow the approach introduced in Ref.~\cite{yang2019deephoyer} where the square-Hoyer term is used since it pushes small interactions (i.e., $J_i<1$) towards zero while respecting larger ones. Mathematically,
\begin{equation}\label{eq:thinning}
    \mathcal{R}_{\text{ET}}(\boldsymbol{J},r,t) = \mathcal{L}(\boldsymbol{J}) +f_{\text{reg}}(r,t)\frac{\left(\sum_i |J_i|\right)^2}{\sum_i J_i^2}
\end{equation}
where $r$ is a regularization constant, $t$ denotes the training epoch, and $f_{\text{reg}}(r,t)$ is an envelope function that schedules the regularizing strength. We find that letting $f_{\text{reg}}(r,t)=r$ for the task in Eq.~(\ref{eq:CHSH}) harms the minimized value reached. Emperically, we find that letting $f_{\text{reg}}(r,t) = re^{-\alpha t/N_{e}}$, where $\alpha=5$ and $N_{e}$ is the total number of epochs trained, works well. This choice of envelope means that the regulatization is scheduled away, steering the learning at early epochs but letting the system minimize $\mathcal{L}$ only at later stages. \par 
Fig.~\ref{fig:Trainability}(c) is a schematic of ET, and Fig.~\ref{fig:Trainability}(e) shows the improved variances of $\partial \mathcal{R}_{\text{ET}}$. Moreover, as the inset shows (see dotted line) the mean gradient improves by several orders of magnitude lifting the BP completely. \par
While ET does not completely zero out the interactions, we observe that for irrelevant qubits $J_i<10^{-5}$ when ET is introduced. Fig.~\ref{fig:Trainability} shows how often weights are effectively thinned away (i.e., made smaller than $10^{-5}$). For this plot, we trained 100 times for 200 epochs each on a $N=4$ QP. We varied the regularization strengths. We see that for $r=0$, it is most likely that all weights are turned on, while for $r=1.0$ one or two irrelevant qubits are often thinned out. For $r=10$ we find that, most often, only one noncontributing qubit is left on. The square-Hoyer term can be easily calculated in a classical computer and therefore does not increase the sample complexity of a QP. \par
In the Sec.~\ref{Sec:Applications}, we explore applications of QPs. However, for these applications no thinning was necessary. Instead, we hope that this section illustrates what to do in case a barren plateau emerges when training a QP. 

\subsection{Using quantum perceptrons as ancillary-based learning}\label{Sec:Ancilla}
\new{Sec.~\ref{Sec:Expressive} showed that quantum perceptrons can perform, in theory, any quantum computation. This claim leveraged the effective Hamitlonian the input qubits experience as a result of their connectivity with the output. Here, we show that the effective Hamiltonian can also be used for learning. We will use this result to apply QPs to problems such as energy measurement in Sec.~\ref{Sec:energy}, and metrology in Sec.~\ref{Sec:metrology}. \par 
Consider again where the output has time-independent fields. Let's call them $f_o^z=0$ and $f^x_o$.
Suppose that $\Delta_o^x$ dominates over the couplings $J_i$. In this case, the input qubits evolve under an effective Hamiltonian 
\begin{align}\label{eq:Effective}
    \heff &= \sigma^x_o\otimes\left(\sum_{ij}\frac{J_iJ_j}{f_o^x}\sigma_i^z\sigma_j^z\right) \notag \\
    &+\mathds{1}_o\otimes\left(\sum_{i=1}^N\sum_{\alpha}f_i^\alpha(t;\bs{\Delta})\sigma_i^\alpha\right) \notag \\
    &= \sigma_o^x\otimes H_{\text{nat}}^{\text{eff}}+\mathds{1}_o\otimes H_{\text{ctr}}^{\text{in}}(t)
\end{align}
Importantly, this evolution is valid if we evolve for times $\tau$ such that $\tau J_{i}J_j/f_o^x\ll 1$. If we choose a time-basis $g_k(t)$ that is sharply peaked around times $t_k=k\tau$ (see Appx.\ref{A:numerics} for details), one can approximate the time evolution of the system by a Trotterized process alternating between the terms $H_{\text{nat}}^{\text{eff}}$ and $H_{\text{ctr}}^{\text{in}}$. That is, up to second order in $\tau$, 
\begin{equation}\label{eq:EffectiveTrotter}
    U_{\text{eff}}(T,\bs{\theta}) = \prod_{l=1}^{L} e^{-i\mathds{1}_o\otimes H_{\text{ctr}}^{\text{in}}(k\tau)}e^{-i\tau \sigma_o^x\otimes H_{\text{nat}}^{\text{eff}}}
\end{equation}
Note that this evolution leaves the output qubit alone if initialized in the states $\ket{\pm}$. However, if initialized in a superposition of them, like $\ket{0}_o = \frac{1}{\sqrt{2}}\left(\ket{+}_o+\ket{-}_o\right),$ we can use the output to measure a quantity on the input. For example, take a single Trotter step $L=1$ with zero $H_{\text{ctr}}^{\text{in}}$. It can readily be shown that the probability of measuring the output qubit in state $|1\rangle_o$ is given by $|\langle \Psi_{\text{in}}|H_{\text{nat}}^{\text{eff}}|\Psi_{\text{in}}\rangle|^2$ where $|\Psi_{\text{in}}\rangle$ is the initial state of the input qubits. Thus, the output qubit's population on the state $|1\rangle_o$ allows us to learn about the energy of the input qubits. This observation will become important for QPs used for energy measurement applications in Sec.~\ref{Sec:energy}.\par 
A second observation about Eq.~(\ref{eq:EffectiveTrotter}), which will become important in Sec.~\ref{Sec:metrology}, is that one can time-reverse the effective evolution by flipping the output qubit from $|+\rangle_o$ to $|-\rangle_o$ (and vice-versa) and by inverting the sign on the time-dependent fields on the input. That is, if one uses the output on state $|+\rangle_o$ and the inputs with fields $\bs{\Delta}$ to effectuate an evolution $U_{\text{eff}}$, changing to $|-\rangle_o$ and $-\bs{\Delta}$ effectuates $U_{\text{eff}}^{\dagger}$.\par 
These observations, make QPs obeying Eq.~(\ref{eq:Effective}) akin to an ancillary-based protocols. Ancilla qubits can be used for operating and learning properties on a several of qubits. For example, they can be used for producing control operations, making nondestructive measurements, detecting errors.}

\section{Applications of quantum perceptrons}\label{Sec:Applications}
We now apply QPs to common quantum computing and quantum machine tasks, specifically to measuring state inner products, entanglement detection, and quantum metrology.\par 

\subsection{Using QPs for measuring state inner products}\label{Sec:overlap}
One important problem in QML is the measurement of kernel functions given a feature map. Suppose we have a data set $\mathcal{D}$ that is a subset of an ambient data set $ \mathcal{X}$. Let $\Psi_{\text{in}}$ be the feature map taking a data point $\bs{x}$ into a quantum state $\ket{\Psi_{\text{in}}(\bs{x})}$. The kernel inherited by the featured map is defined as $\kappa(\bs{x},\bs{y}) \equiv |\langle\Psi_{\text{in}}(\bs{x})|\Psi_{\text{in}}(\bs{y})\rangle|^2$~\cite{schuld2019quantum,huang2021power}. The Representer Theorem~\cite{scholkopf2001generalized} tells us that for a given machine learning task (a data set, a loss function, and a set of labels), the function $g^{*}$ minimizing the loss can be written in terms of the kernel: $g^*(\bs{x}) = \sum_{\bs{y}}\alpha_{\bs{y}}\kappa(\bs{x}, \bs{y})$. It is commonly argued that quantum kernels are more expressive than those obtained by classical. Thus, a central goal of QML is to measure such kernels efficiently. Note that measuring these kernels amounts to measuring the inner product between two states. Inner products can be measured using the SWAP-test~\cite{buhrman2001quantum} in $\mathcal{O}(1)$ operations, assuming one can do control-SWAP gates in parallel. This section shows that such paralleling of control-SWAP gates is natural to a QP setup.\par
Let us first show how one could measure the inner product between two one-qubit states using a QP. Our QP has two input qubits $i=1,2$ encoding the data $\bs{x}_1$ and $\bs{x}_2$ in $\mathcal{D}$ into the one-qubit states $|\Psi_{\text{in}}(\bs{x}_1)\rangle$ and $\ket{\Psi_{\text{in}}(\bs{x}_2)}$. Explicitly, 
\begin{align*}
    |\Psi_{\text{in}}(\bs{x}_1)\rangle_1 &= a_0(\bs{x}_1)\ket{0}_1+a_1(\bs{x}_1)\ket{1}_1\\
    |\Psi^{*}_{\text{in}}(\bs{x}_2)\rangle_2 &= a^*_0(\bs{x}_2)\ket{0}_2+a^*_1(\bs{x}_2)\ket{1}_2.
\end{align*}
Here $|\Psi_{\text{in}}(\bs{x}_1)\rangle$ lives on the first input qubit, and $|\Psi^{*}_{\text{in}}(\bs{x}_2)\rangle$ on the second input qubit. Suppose that these two input qubits are connected to an output qubit via a Hamiltonian with $J_1=-J_2=J$, $f^z_o=0$, and $f^x_o=\pi$ with $J\gg f^x_o$. Then, the system evolves to 
\begin{align}
    &\sum_{ij} e^{-i(J((-1)^i-(-1)^j)\sigma^z_o+\Delta^x_o\sigma_o^x)}a_i(\bs{x}_1)a^{*}_j(\bs{x}_2)|ij0\rangle \notag \\
    &\approx\sum_{i=j}a_i(\bs{x}_1)a^*_j(\bs{x}_2)|ij1\rangle \notag\\
    &+\sum_{i\neq j}e^{-i(\pm 2J\sigma^z_o)}a_i(\bs{x}_1)a^{*}_j(\bs{x}_2)|ij0\rangle
\end{align}
Thus, the probability of the output qubit being on state $\ket{1}$ is precisely $P(1) = |\sum_ia_i(\bs{x}_1)a^{*}_i(\bs{x}_2)|^2 = |\langle \Psi_{\text{in}}(\bs{x}_1)|\Psi_{\text{in}}(\bs{x}_2)\rangle|^2$, which is the kernel $\kappa(\bs{x}_1,\bs{x}_2)$. Note that this results from the condition $J\gg f^x_o$ since only when $i=j$ the coherent drive $f^x_o\sigma_o^x$ can significantly act on the output qubit. In the atomic physics community this condition is known as the rotating wave approximation.\par 
One can generalize to larger input sizes straightforwardly. Suppose that state $|\Psi_{\text{in}}(\bs{x}_1)\rangle$ now extends over $n$ qubits $i=1,...,n$ and that state $|\Psi^{*}_{\text{in}}(\bs{x}_2)\rangle$ extends over also $n$ qubits $i=n+1,...,2n$. If we use $J_i = -J_{i+n}=J$,  $J\gg f^x_o=\pi$ with $f^z_o=0$, and all other fields zero, then the probability of the output being on state $\ket{1}$ is again $P(1) = |\langle \Psi_{\text{in}}(\bs{x}_1)|\Psi_{\text{in}}(\bs{x}_2)\rangle|^2$. \par 
Our approach employs the dynamics of a QP to realize a SWAP test in a constant number of operations. 
Note that $H_P$ treats all the input qubits as neighbors to the output qubit. Thus, implementing QPs relies on high-connectivity (i.e. long-range direct interaction) or programmable platforms. In Sec.~\ref{Sec:experimental}, we propose implementing QPs in several platforms, such as superconducting qubits~\cite{IBM_eagle}, cold-atom arrays~\cite{singh2022dual,browaeys2016experimental, de2019observation,bluvstein2112quantum,PhysRevLett.128.050502, cody2022transport,rui2020} and nitrogen-vacancy centers in diamond~\cite{bradley2019ten,PhysRevX.10.031002}.\par 
The scheme presented above is a particular case of the Hadamard test~\cite{aharonov2006polynomial}. The Hadamard test uses an ancillary to measure $\bra{\psi}U\ket{\psi}$ where $U$ is some unitary and $\ket{\psi}$ a state. Including pulses in the input of a QP realizes the Hadamard test since pulsing effectuates any unitary $U$. The output qubit can then be used as the ancillary to read the desired expectation value. \par

\subsection{Using QPs for energy measurement and entanglement detection}\label{Sec:energy}
\begin{figure}
    \centering
    \includegraphics[width=0.9\linewidth]{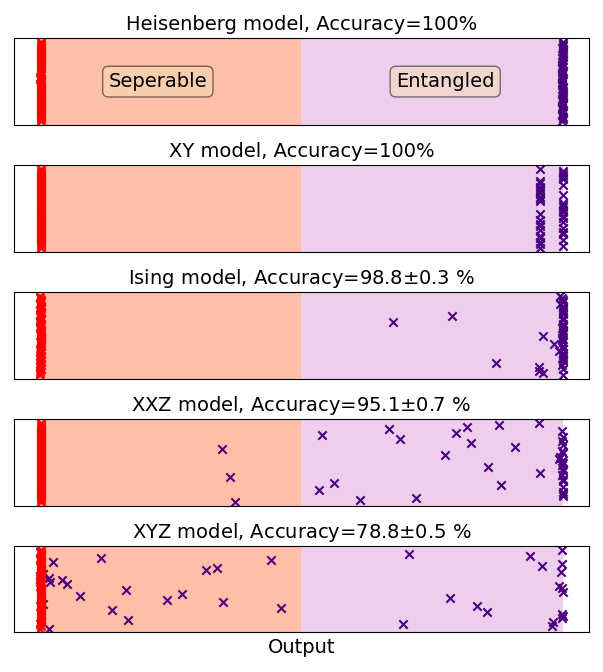}
    \caption{Results of using QPs for entanglement detection. The purple crosses correspond to entangled states, while the red crosses correspond to separable states. The purple region is where the machine is trained to label entangled. The top panel shows the classification results on a test set made out of ground states of 1D Heisenberg models and their mean-field approximations. See the main text for a thorough discussion of the data set. For this task, $N=4$ and $L=4$ sufficed to give unit accuracy on the test set. The rest of the panels show the performance of the QPs on test sets consisting of ground states of other spin models of the form in Eq.~(\ref{eq:Heisenbergs}) but are not included in the training set. The XY model corresponds to $J_{z}=h=0$ and $J_x=J_y=J$, the Ising model to $J_z=J$, $J_x=J_y=0$ and $h=0.5$, the XXZ model for which $h=0$ and $J_x=J_y$ and $J_z\neq 0$, and the XYZ model for which $h=0$ and all couplings are independent. We see that the learned witness of the QPs can detect entanglement in different families of states. Moreover, the QP only labels separable states as separable. This correct labeling is due to the sufficiency of the witness.}
    \label{fig:entanglement_witness}
\end{figure}

As discussed in Sec.~\ref{Sec:Ancilla}, QPs operating under the effective Hamiltonian in the inputs in Eq.~(\ref{eq:EffectiveTrotter}) can be used for energy measurement. In fact, with the time-dependent fields on the input, and $H_{\text{nat}}^{\text{eff}} = \sum_{ij}\frac{J_iJ_j}{f^x_o}\sigma_i^z\sigma_j^z$, one could rewrite $U_{\text{eff}}$ as $e^{-i\sigma_o^x\otimes \tilde{H}_{\text{eff}}}$ where the effective Hamiltonian is a generalized Heisenberg model~\cite{PhysRevX.10.031002}. The result in Eq.~(\ref{eq:expectationEnergy}) tells us that a QP can measure the energy of a state under a Hamiltonian $\tilde{H}_{\text{eff}}$. That is, if we start the output qubit in $|0\rangle_o$, the output qubit's expectation of $\sigma_o^z$ is 
\begin{equation}\label{eq:expectationEnergy}
    \langle \sigma^z_o(\tau)\rangle\approx 2\tau \bra{\Psi_{\text{in}}}\tilde{H}_{\text{eff}}\ket{\Psi_{\text{in}}}.
\end{equation}
where $|\Psi_{\text{in}}\rangle$ is the initial state of the input qubits. \par 
We highlight that measuring energy as in Eq.~(\ref{eq:expectationEnergy}) relies on measuring a single one-qubit observable as a function of time. This is far less expensive than traditional circuit-based variational quantum algorithms~\cite{peruzzo2014variational, liu2021representation, liu2022analytic} methods where $\mathcal{O}(N)$ two-point correlators need to be measured directly on a the many-body input system. Thus, our QP can measure the energy of a generalize Heisenberg Hamiltonians with enhanced precision and efficiency. \par
We can use energy measurements under a many-body Hamiltonian for entanglement detection~\cite{dowling2004energy, toth2005entanglement, sperling2013multipartite}. For certain Hamiltonians, the energy of its lowest-energy eigenvalues is lower than that of any separable state~\cite{sperling2013multipartite}. To explain this, let us define the energy gap of a generic Hamiltonian $\hat{H}$ as 
\begin{equation}
    G = E_{\text{sep}}-E_0
\end{equation}
where $E_0$ is the ground state energy of $\hat{H}$, and $E_{\text{sep}}$ is the lowest energy of a separable state under $\hat{H}$. The entanglement gap tells us about the properties of the Hamiltonian. For example, $G>0$ if no state in the ground state manifold is separable. Every Hamiltonian with a positive entanglement gap defines an entanglement witness operator
\begin{equation}\label{eq:witness}
    W = \hat{H}-E_{\text{sep}}\mathds{1}
\end{equation}
since measuring $\langle\Psi|W|\Psi\rangle<0$ on a state $|\Psi\rangle$ suffices to show that $|\Psi\rangle$ is entangled (i.e. not separable). We seek Hamiltonians with large entanglement gaps since these can detect entanglement even if the states have local imperfections or are at finite temperature~\cite{dowling2004energy}.\par 
Consider the generalized Heisenberg Hamiltonians, which are of the form 
\begin{align}\label{eq:Heisenbergs}
    H(J_{x,y,z},h) &= \sum_{i}\left(J_x\sigma_i^x\sigma_{i+1}^x+J_y\sigma_i^y\sigma_{i+1}^y+J_z\sigma_i^z\sigma_{i+1}^z\right)\notag\\
    &+h\sum_{i}\sigma_i^x.
\end{align}
The Heisenberg model $H_{\text{Heis}}$ is the case where $J_x=J_y=J_z=J$ and $h=0$. In the case of the $H_{\text{Heis}}$ the entanglement gap is $G \approx 0.77 N$ so that if it is used for witnessing the quality of the witness improves linearly in $N$~\cite{toth2005entanglement}. \par
We now use a QP to produce an entanglement witness out of $\tilde{H}_{\text{eff}}$. If Eq.~(\ref{eq:expectationEnergy}) has a negative value for entangled states and a positive value for separable ones, $\tilde{H}_{\text{eff}}$ will have a nonzero entanglement gap and thus operate as a witness. Measuring the output of a QP then reveals the energy under this Hamiltonian. This measurement indicates the value of a potential witness up to a constant factor.  \par

In particular, we set $N=4$ and proceed to generate a data set $\mathcal{D}$ consisting of two types of states. This first type in $\mathcal{D}$ are ground states $|\text{GS}_J\rangle$ of 1D Heisenberg Hamiltonians with coupling-strength $J$ sampled uniformly in the interval $[-1,1]$. The second state type is the set of mean-field approximations $|\text{MF}_{J}\rangle$ of each ground state $|\text{GS}_J\rangle$,  which can be written as the separable states $|\text{MF}_{J}\rangle = \bigotimes_{i=1}^N\left(\cos\theta_i\ket{0}+e^{i\phi_i}\sin\theta_i\ket{1}\right)$. The parameters $\{\theta_i,\phi_i\}$ are searched using gradient descent to minimize the energy $E_{\text{sep}}=\langle \text{MF}_J|H_{\text{Heis}}|\text{MF}_{J}\rangle$. Our data set consists of 200 ground states labeled with a value ``$-1$'' for entangled and 200 mean-field states labeled with a ``$1$'' for separable. Thus, $\mathcal{D} = \{ (|\text{GS}_J\rangle, -1), (|\text{MF}_J\rangle,1)\}_{J}$. Each state $|\psi\rangle$ in the data set is passed through a series of QPs with $\tau=0.01,0.02,\hdots,0.1$ and fixed $f^x_0=50$. We note that this setup is not completely within the requirement $f^x_0\tau\ll 1$ that we used to derive Eq.~(\ref{eq:EffectiveTrotter}), but that it serves as a physically-inspiring starting point for gradient descend to operate optimize the variational parameters given that Eqs.~(\ref{eq:expectationEnergy}) and (\ref{eq:witness}) shows that a QP can theoretically be used for entanglement witnessing. \par 
For each QP, we measure Eq.~(\ref{eq:expectationEnergy}) and collect a vector $\bs{r}(\psi) = (\langle\sigma^z(0.01)\rangle, \hdots, \langle\sigma^z(0.1)\rangle, 1)$. The last entry of $\bs{r}(\psi)$ is a bias term. An output is produced by post-processing the measurements to obtain
\begin{equation}
\tilde{y}(\psi) = \text{softsign}(\bs{w}\cdot\bs{r}(\psi))
\end{equation} 
where $\bs{w}$ is a vector of optimization parameters and $\text{softsign}(x) = x/(1+|x|)$. We calculate the square-loss in Eq.~(\ref{eq:LossMultiple}) by summing over the states $\psi\in \mathcal{D}$. All the QPs have the same parameters, and only $\tau$ differs. The loss minimizes over the QP parameters and the vector $\bs{w}$. We choose to measure $\langle\sigma_o^z(\tau)\rangle$ for a variety of $\tau$ to give the post-processing more information to pin-point the energy of the input state. For a vector $\boldsymbol{r}$ of $n$ entries, and a desired additive error $\varepsilon$ we note that obtaining $\tilde{y}$ requires running the time evolution a total of $\mathcal{O}(n/\varepsilon^2)$ times; a repetition rate independent of system size but depending on the hyper-parameters $n$ and the desired accuracy.
Fig.~\ref{fig:entanglement_witness} shows the results of using QPs for entanglement detection. The top panel shows the values of $\tilde{y}_{\bs{\theta}}$ for test states obtained from a Heisenberg model. We measure the accuracy as the percentage of test states that satisfy $|y(\psi)-\tilde{y}_{\bs{\theta}}(\psi)|<1.0$. We note that $L=4$ was sufficient to show unity accuracy on the Heisenberg test set. \par 

field approximations) of other 1D spin models not included in the training set. In particular, we tested against the XY model for which $J_{z}=h=0$ and $J_x=J_y=J$~\cite{lieb1961two}, the Ising model for which $J_z=J$, $J_x=J_y=0$ and $h=0.5$~\cite{stinchcombe1973ising}, the XXZ model for which $h=0$ and $J_x=J_y$ and $J_z\neq 0$, and the XYZ model for which $h=0$ and all couplings are independent~\cite{sachdev1999quantum}. All nonzero couplings $J_{x,y,z}$ were sampled uniformly on the interval $[-1,1]$. The test sets consisted of 200 states each. Note that our QPs learned to classify all separable states for all models, while only some entangled states were classified correctly. This correct labeling is because entanglement witnessing is a sufficient and not necessary test of entanglement. We choose to train the QP using the ground states of Heisenberg Hamiltonians since we know that the Heisenberg model has a large energy gap suitable for witnessing. However, we could have also chosen to train using other ground states. Fig.~\ref{fig:entanglement_witness} suggests that QPs trained on the Heisenberg model result in a powerful witness that can detect entanglement in other families of states not seeing during training. \par 
The results in this section showcase the power of QPs for entanglement detection, a critical task in quantum computing and information. Thus, this section motivates using QPs to measure global dynamical quantities of quantum many-body systems.\par 

\subsection{Time-reversal and metrology using QPs}\label{Sec:metrology}
\begin{figure}[t]
\begin{center}
\includegraphics[width=\linewidth]{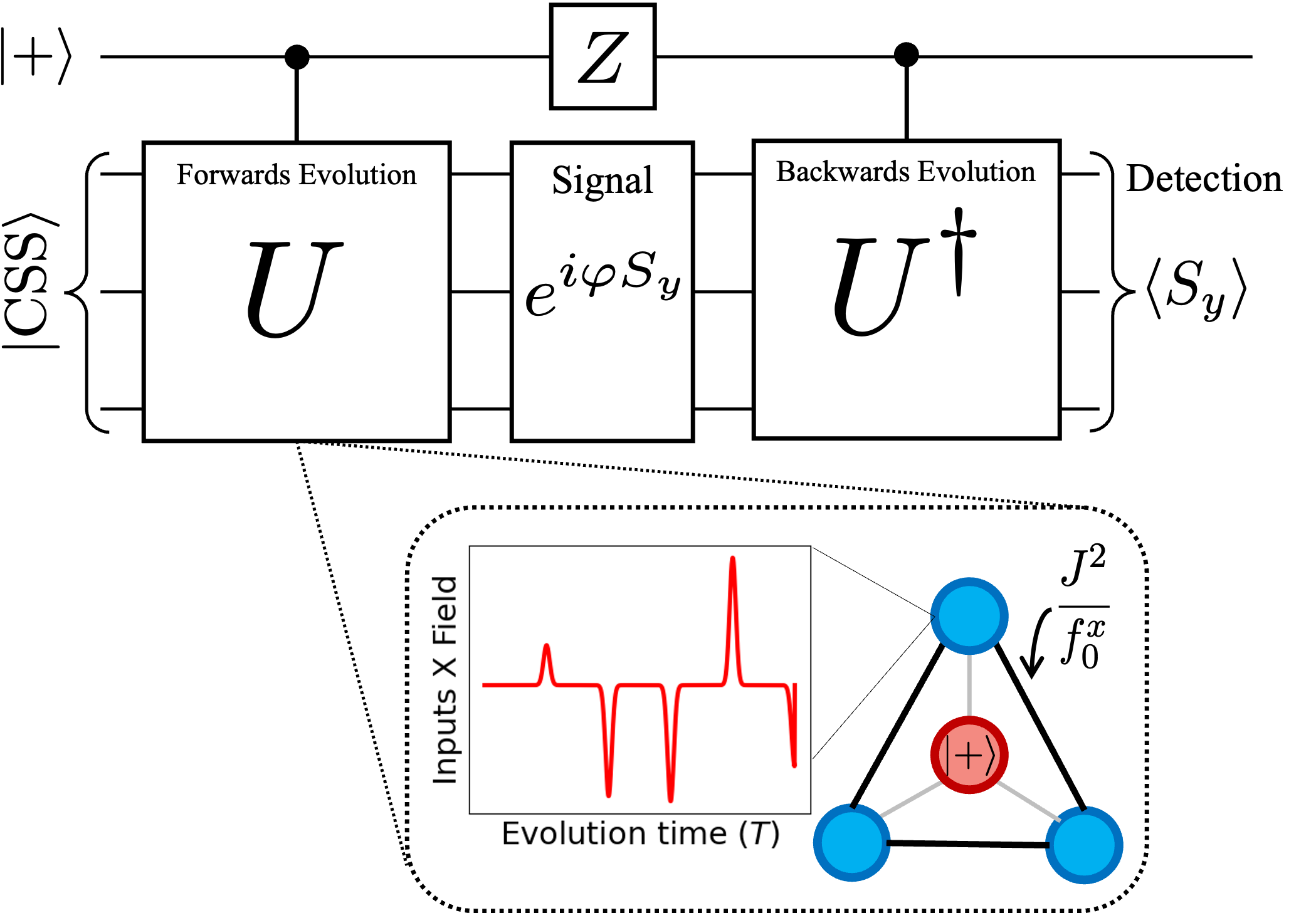}
\end{center}
\caption{Using a quantum perceptron (QP) for time-reversal and applying that for metrology. In other words, using a QP to detect a signal $\varphi$ on the input qubits. The input is evolved from a coherent spin state (CSS) $|+\rangle^{\otimes N}$ to a spin-squeezed state via the effective Hamiltonian generated by the output qubit. This effective Hamiltonian is  $H_{\text{nat}}^{\text{eff}}\propto S^2_z$, which generates squeezing useful for metrology. Here $S_z$ is the total spin of the input qubits. We add global fields on the inputs to aid with squeezing. After the signal operates on the input, the output node changes from $|+\rangle_o$ to $|-\rangle_o$, effectively time-reversing the effective Hamiltonian. The fields are sign reverse by hand. We measure $\langle S_y\rangle$ on the input. While for $\varphi=0$ the state returns to the CSS with $\langle S_y\rangle=0$, for small $\varphi$ we obtain $\langle S_y\rangle=m\varphi N/2$ where $m$ is the signal amplification.}\label{fig:metrology_circuit}
\end{figure}

\begin{figure}[t]
\begin{center}
\includegraphics[width=0.9\linewidth]{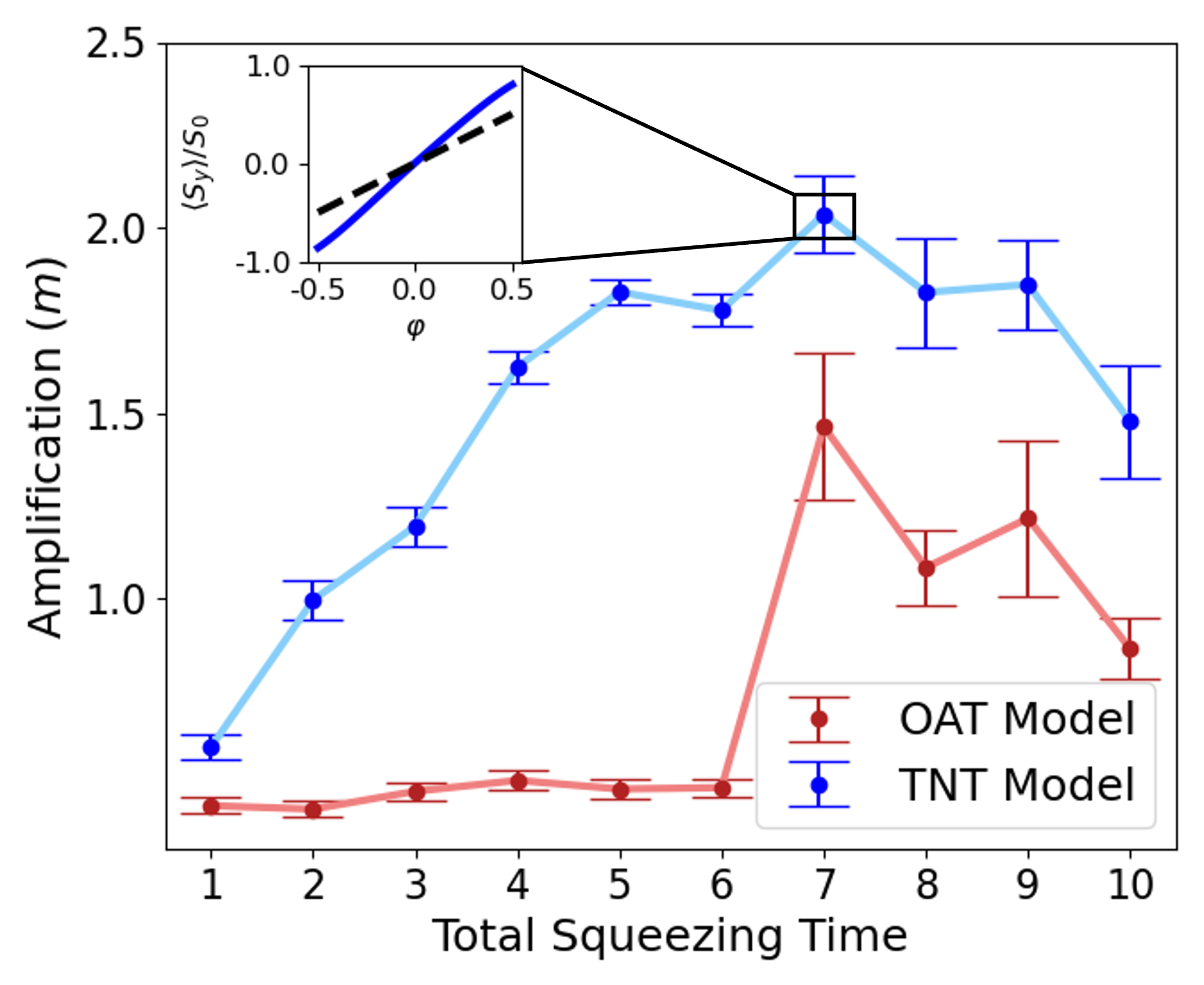}
\end{center}
\caption{We show a variational time-reversal protocol for metrology on an $N=10$ quantum perceptron. We obtain the amplification $m$ by measuring the input's total spin $\langle S_y\rangle = m\varphi S_0$ where $S_0=N/2$ as in Fig.~\ref{fig:metrology_circuit}. We plot $m$ as a function of the number the total evolution time ($T$). We present the results for the OAT model without global rotations along the x-axis (red lower curve) and the TNT model with global rotations along the x-axis (blue higher curve). As expected, the TNT model achieves a higher amplification $m$. The maximum $m$ is $2.01$, corresponding to a metrological gain of $3.03$ dB. Note that the gain is smaller than the maximum for small and large values of $T$. For those values of $T$, our Hamiltonian produces a state under and over-squeezed, respectively. The inset shows a typical curve of the obtained $\langle S_y\rangle$ as a function of $\varphi$ for the TNT model at $T=7$. We see that the $\langle S_y\rangle$ closely fits a linear function with a slope $m>1$. Note that $m<1$ is worse than doing nothing at all. We include the $m=1$ line as a reference (dotted black line). Our maximum metrological gain of $3.03$ dB is about 2 dB below the best-achieved gain for $N\sim 10$ using maximally-entangled states in Rydberg tweezer arrays~\cite{omran2019generation}. However, QPs Hamiltonian evolution is less resource intensive as it relies on the system's dynamics and not on adiabatic state preparation.}
\label{fig:metrology}
\end{figure}

From Eq.~(\ref{eq:EffectiveTrotter}), we see that flipping the output qubit from $|+\rangle_o$ to $|-\rangle_o$ changes the sign of the effective Hamiltonian $H_{\text{nat}}^{\text{eff}}$ on the input qubits. A corollary to this is that a QP can be used to perform time-reversed unitary evolution on the input qubits. This time-reversal capability is an example of a well-known, ancillary-assisted operation similar to those in cold atoms in cavities~\cite{PhysRevLett.104.073602,swingle2016measuring, RevModPhys.89.035002}. With this observation, we can use the output qubit to steer and measure the input similar to in nitrogen-vacancy experiments~\cite{PhysRevLett.106.140502}. \par
Additionally, the effective Hamiltonian in Eq.~(\ref{eq:EffectiveTrotter}) has the form (for uniform couplings)
\begin{equation}\label{eq:S2}
    H_{\text{nat}}^{\text{eff}} = \frac{J^2}{f^x_o}\sum_{ij}\sigma^z_i\sigma^z_j \propto S^2_z.
\end{equation}
where $S_z = \frac{1}{2}\sum_{i=1}^{N}\sigma_i^z$ is the total input spin in the z-axis. This non-linear Hamiltonian is known to create highly entangled states by spin-squeezing initially separable states~\cite{PhysRevLett.104.073602,swingle2016measuring, colombo2022time}. \par 
Highly entangled states are sensitive to small perturbations. Time-reversing entangling operations can help us measure the perturbation. The recently proposed signal amplification time-reversal interactions protocol (SATIN) illustrates this idea. In SATIN, a simple initial state is evolved to be highly entangled via a Hamiltonian of the form $S_z^2$ --- called the one-axis twisting model (OAT)~\cite{PhysRevA.47.5138} --- before interacting with a signal to be measured $\varphi$. Subsequently, the entangled state is time-reversed by evolving under $-S_z^2$. The final state is displaced from the original one, and under certain conditions, the signal $\varphi$ is amplified so that it can be measured~\cite{colombo2022time}.
Concretely, as in Ref.~\cite{colombo2022time}, the perturbation is a global rotation along the y-axis $e^{-i\varphi S_y}$ where $S_y =\frac{1}{2}\sum_{i=1}^N\sigma_i^y$. The initial state is the coherent spin state (CSS) $|+\rangle^{\otimes N}$ for which $\langle S_y\rangle = 0$. After the time-reversal, however, if $\varphi\neq 0$, the state does not return to the CSS and instead~\cite{colombo2022time}
\begin{equation}
    \langle S_y\rangle = m \varphi N/2.
\end{equation}
Here, $m\geq 1$ is called the amplification of the perturbation $\varphi$, and we denote $S_0=N/2$.\par
In a QP we achieve time-reversal by changing the sign of all the time-dependent controls and switching the state of the output qubit from $|+\rangle_o$ to $|-\rangle_o$ via applying the operator $Z=\ket{+}\bra{-}+\text{h.c.}$. Unlike the original SATIN proposal, a QP performs single-qubit operation as well. As explained in Appx.~\ref{A:numerics}, we choose the fields to be only on for $L$ small windows of time (i.e., we Trotterize the interactions and single-qubit controls). In between the nonzero fields, the system evolves under $H_{\text{nat}}^{\text{eff}}$ for a time $\tau$ such that $\tau J_{i}J_{j}/f^x_o\ll 1$. The total evolution time is $T=L\tau$. Thus, our QP repeatedly twists the input qubits followed by a quick rotation of them along the $x$-axis. This process is known in the literature as the ``twist-and-turn model'' (TNT)~\cite{liu2011spin,PhysRevA.92.023603}. We expect the TNT model to produce higher amplification than the OAT model~\cite{PhysRevA.97.053618}. By optimizing over the interactions and control fields, QPs realize a variational SATIN protocol. Fig.~\ref{fig:metrology_circuit} shows a schematic of the procedure.\par 

Before diving into our application's results, we review the sensitivity to the signal ratios in different measurement protocols. For the typical case of using Ramsey interferometry~\cite{ramsey1950molecular} on $N$ independent qubits to measure $\varphi$, the sensitivity of the probe is limited to $(\Delta \varphi)_{SQL} = 1/\sqrt{N}$. This result is known as the standard quantum limit (SQL)~\cite{danilishin2012quantum}. On the other hand, the precision can improve to $1/N$ by using a system of $N$ highly entangled qubits, a result known as the Heisenberg Limit~\cite{huelga1997improvement, leibfried2004toward, appel2009mesoscopic}. This limit, however, requires the production of a maximally entangled state, a feat only doable for a small number of qubits with significant experimental burdens~\cite{song201710, omran2019generation}. A more modest approach aims to achieve a Heisenberg scaling on the precision of $b/N$ with $b>1$~\cite{saffman2009spin}, which SATIN can accomplish.  \par 
The sensitivity of SATIN is approximated by $\delta \varphi = 1/(m\sqrt{N})$. As usual, we define the metrological gain of SATIN as 
\begin{equation}
    \mathcal{G} = \frac{(\Delta \varphi)^2_{SQL}}{\delta\varphi^2} = m^2.
\end{equation}
Fig.~\ref{fig:metrology} shows the results of the variational SATIN protocol for $N=10$. We plot the amplification $m$ as a function of $L$ and the average results of 50 runs. Error bars represent one standard deviation from the mean. The OAT and TNT models achieve a maximum $m$ at $T=7$. This maximum is consistent with the results in Ref.~\cite{colombo2022time} since the state before the perturbation is barely squeezed for small $L$ (i.e., small total time). Likewise, for large total time, the squeezing is so large that it washes out the signal. However, the highest amplification $m=2.01$ is achieved with the TNT model leading to $\mathcal{G} = 3.01$ dB.\par 
Interestingly, our protocol already outpaces SQL at $N=10$, contrary to the predictions in Ref.~\cite{colombo2022time}. We highlight that only Rydberg tweezer arrays have achieved a metrological gain $m>1$ at $N\sim 10$~\cite{omran2019generation}. However, the QPs are less resource intensive that the approach in Ref.~\cite{omran2019generation} which uses maximally entangled states. Thus, we expect that variational SATIN protocol on QPs improve on metrology protocols that use time-reversal operations. QPs' ability to steer the input qubits with a nonlinear Hamiltonian condition on the output qubit's state is helpful for metrology applications relevant to quantum information and processing. \par 

\section{Potential experimental implementations of QPs}\label{Sec:experimental}
\begin{figure}[b]
\begin{center}
\includegraphics[width=0.80\linewidth]{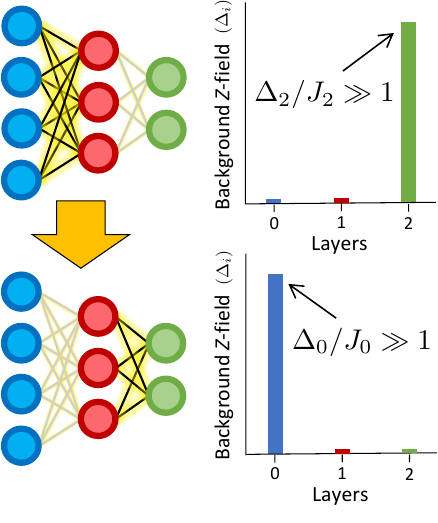}
\caption{Experimental proposal for QNC architectures composed of several quantum perceptrons. Perceptrons can be ''stacked'' together to form a single-layer neural network. In architectures without coherent qubit re-arrangement, several layers can be implemented by freezing out all but one layer at a time. This can be realized by implementing a background layer-dependent $Z$-field ($\Delta_i$) such that in inactive layers $\Delta_i\gg J_i$, where $J_i$ is the typical interaction strength of each inactive layer. If no layers are frozen-out, one recovers a quantum reservoir computer.}
\label{fig:Experimental}
\end{center}
\end{figure}
Among the greatest motivations for pursuing this manuscript's manifestation of quantum perceptrons is that $H_P$ is an interaction-rich Hamiltonian suitable for near-term devices. In some cases, $H_P$ is already available. \par 
QPs consist of only two types of gates: analog evolution under $H_P$ and single-qubit rotations. Thus, QPs can be implemented by several quantum computing and quantum simulation platforms in the near term. This provides a straightforward protocol to implement the QP in gate based highly controllable and pulse-equipped quantum hardware such as superconducting qubit~\cite{IBM_eagle, alexander2020qiskit}.  \par
However, the QP can also be realized in other analog quantum hardware, and they can be used as the building blocks of more intricate QNC architectures. For instance, QPs could be realized with Rydberg atom arrays that contain two atomic species~\cite{singh2022dual}, with one species being used for measurement and manipulation of the other. Currently, however, this platform can only perform global rotations and restricted site-dependent addressing, which may decrease the expressivity of the implementation but can still realize tasks such as the one in Sec.~\ref{Sec:metrology}. A variation on this platform with two Rydberg states and one atomic species can also be used to implement QPs. For example, experiments in~\cite{browaeys2016experimental, de2019observation} ensembles of Rubidium-87 Rydberg atoms are used where two Rydberg excitations ($60S$ and $60P$) are used so that atoms in the $60S$ state interact strongly with atoms in the $60P$ state while inter-state interactions are negligible. For example, a QP could be realized by encoding all inputs in the $60S$ state and retrieving the output in the $60P$ state. \par 
Recent developments in cold atom arrays' re-configurable geometries~\cite{bluvstein2112quantum} and movable trapped ions~\cite{PhysRevLett.128.050502, cody2022transport} can also implement QPs. In both platforms, atoms can be moved around to create effective interactions. As a result, the input qubit of a QP can be made to interact with the output qubits one at a time, avoiding input cross-talk. The separation between the input and output qubits also allows for flexible implementation of single-qubit control. \par
Atomic arrays with wavelength spacing, such as those generated by optical lattices, are also promising platforms for QPs. The strong, short-range dipole-dipole interactions of these structures endow them with rich, tuneable atom-atom interactions~\cite{masson2020}. In the case of multiple atomic species or states, these systems' interaction length and coherence times can be expanded through array-mediated atom-atom dynamics~\cite{patti2021controlling}, and analogous implementations have also been proposed for atomic dimers~\cite{castells2021}. Two-dimensional lattices of wavelength spacing have particular potential due to their ability to interact deterministically with a single photon~\cite{shahmoon2017} and their recent experimental implementations~\cite{rui2020}. 
\par
Passing from atomic quantum computers to solid-state platforms, we note color centers in diamonds as a promising system for implementing QPs. For instance, $H_P$ resembles the so-called ``star model'' used to describe how an electronic spin in a nitrogen-vacancy center in a diamond interacts with neighboring nuclear spins~\cite{bradley2019ten}. Thanks to recent developments in single-qubit rotation sequences~\cite{PhysRevX.10.031002,PRXQuantum.3.020303,geier2021floquet}, single-qubit control is attainable within the near feature. However, the downside of this platform is the lack of tunability of the coupling constants $J_{i}$ in Eq.~(\ref{eq:Hp}).\par 
Moreover, all of these platforms could theoretically enable the building of more complex QNC architectures or networks. For coherently re-configurable platforms (those with qubit transport), this can be by programming the qubit's movement to emulate the network's architectures; a protocol that would rely on finding efficient methods to move several qubits at once to accommodate the network's connectivity within the coherence times of the platform. For all other platforms hereby proposed, neural networks could be implemented as in Fig.~\ref{fig:Experimental}. First staking QPs together forms a graph as in a classical artificial neural network. Secondly, a layer-dependent background $Z$-field can then be used to freeze out layers such that information propagates from layer-to-layer as in the classical case. If no such ordered propagation is desired, one could leave all qubits active thereby recovering a quantum reservoir computer. The exact implementation of these QNC architecture is an area of active research left for future work. We highlight that it will depend on the exact target platform and the task at hand.

\section{Conclusion}\label{Sec:Conclusion}
In this manuscript, we introduce quantum perceptrons (QPs) based on the unitary dynamics of spin systems. We show that QPs are helpful for common quantum machine learning problems, such as measuring the inner product between two states. We prove that QPs can approximate classical perceptrons and thus approximate multi-variable functions using multiple QPs. The approximation is noise insensitive to changes in the input. Therefore, architectures built out of QPs promise to be noise-resilient. \framing{We also show that QPs can be used to realize any quantum computation, establishing that a single QP offers far more computational complexity than its classical counterpart. Our proof, however, is not to be used to claim that QPs are advantageous over other quantum computing schemes.} We also discuss using QPs for machine learning tasks by optimizing the couplings and local time-dependent fields. We exemplified that QPs may experience barren plateaus which can problematize their training, and proposed a technique to successfully mitigate them.\par 
The studies in this manuscript address a question that has recently been debated as a fundamental one for QML: "what are good building blocks for quantum [machine learning] models?"~\cite{schuld2022quantum}. Out QPs answer this question with a simple and modular architecture. This architecture is inspired by quantum simulation tools which may enable QPs to be scalable soon. Our QPs are also insensitive to noise in the input states (Sec.~\ref{Sec:Recovery}), and their "non-classicality" clearly stems from single-qubit pulses, as seen in Sec.~\ref{Sec:Expressive}. In Sec.~\ref{Sec:experimental} we discussed potential ways to implement QPs and build quantum neuromorphic neural networks using them. \par 
Moreover, QPs are especially well-suited for macroscopic observable measurements of the input qubits using the output qubit as a probe. In particular, we showed that QPs could measure energies of many spin models (Sec.~\ref{Sec:energy}). Thus, QPs measure quantities accurately and efficiently that would otherwise be hard to compute directly on a many-body system. We argue that QPs are an essential step towards scalable quantum machine learning applications on near-term devices for the reasons highlighted above.\par
Nonetheless, a further work is needed to discover the full potential of QPs. Directly from this work, investigations into the trade-off between the fields' complexity and higher number of qubits are of great interest. These studies can be done by either stacking QPs or connecting the inputs at specific intermediate steps. Large-scale QP implementations may also benefit from gradient-free training schemes such as those recently developed for other analog algorithms~\cite{leng2022differentiable}. Another direction of interest is using QPs for distributed computing, where a quantum computer prepares a particular state, and a QP is used to encode macroscopic quantities of the state, which can then be transported to do further computation~\cite{cuomo2020towards}, or using the restricted connectivity of QPs as blocks in larger computation~\cite{beals2013efficient}. QPs may also be incorporated as subroutines of other variational quantum circuits along the lines of the measurement-based perceptrons in~\cite{gili2022introducing}. Recent developments in dynamically re-configurable computing geometries may realize this process~\cite{bluvstein2112quantum,PhysRevLett.128.050502, cody2022transport}. While in Sec.~\ref{Sec:Applications} our algorithms used short-time dynamics, the type of problems that long-term dynamics can solve remains an open question. These studies will likely benefit from an analytical understanding of statistical ensembles of many-body systems. \framing{While we do not claim that QPs are advantegeous over other quantum computing schemes, we encourage our community to use QPs when it fits their goals and particular experimental platforms.} 

\begin{acknowledgments}
The authors thank Di Luo, Hong-ye Hu, and Pengyuan Bill Zhai for insightful discussion, Simone Colombo for pointing out how a QP can be used for time-reversal, and Bailey Renger, an intern researcher at Harvard University during the Spring of 2022 who tested the simulations in Sec.~\ref{Sec:Recovery}. RAB acknowledges support from National Science Foundation (NSF) Graduate Research Fellowship under Grant No. DGE1745303. XG kindly acknowledges support from the Harvard-MPQ Center for Quantum Optics, the Templeton Religion Trust (TRT) under Grant No. 0159, and by the Army Research Office under Grant No. W911NF1910302 and the Multidisciplinary University Research Initiative (MURI) under Grant No. W911NF-20-1-0082.SFY acknowledges funding from NSF Q-IDEAS HDR Institute, the Q-SeNSE QCLI and the CUA PFC, and the Air Force Office for Scientific Research (AFOSR).
\end{acknowledgments}

\bibliographystyle{unsrtnat}
\bibliography{bibliography}

\onecolumngrid
\appendix

\section{Supplementary Calculations for Recovering the Classical Perceptron}\label{A:Pulsing}
Let us initialize the system in a state $|\Psi_{\text{in}}\rangle = \sum_{\bs{s}}\psi_{\bs{s}}\ket{\bs{s}}\ket{0}_o$ where $\ket{\bs{s}}$ is an eigenstate for each Pauli-Z on the input (i.e., $\sigma_i^z\ket{\bs{s}} = s_i\ket{\bs{s}}$ with $s_i=\pm 1$). According to Eq.~(\ref{eq:Hp}) for the case that all the inputs experience zero fields, and the output experience non-zero constant fields along the $x$ and $z$ axis, the system evolves to 
\begin{align}
    &\sum_{\bs{s}}\psi_{\bs{s}}\ket{\bs{s}} \left[\cos{\omega(\bs{s})}-i\frac{h(\bs{s})}{\omega(\bs{s})}\sin{\omega(\bs{s})}\right]\ket{0}_o\notag\\
    +&\sum_{\bs{s}}\psi_{\bs{s}}\ket{\bs{s}}i\frac{f^x_o}{\omega(\bs{s})}\sin{\omega(\bs{s})}\ket{1}_o
\end{align}
where 
\begin{align}
    h(\bs{s})&=\sum_{i=1}^{N}J_is_i-f^z_o, \\
    \omega(\bs{s}) &= \sqrt{h(\bs{s})^2+\left(f^x_o\right)^2}.
\end{align}
To see this, note that for any one of the basis states $\ket{\bs{s}}\ket{0}_o = \ket{s_1,...,s_N}\ket{0}_o$, we have that Eq.~(\ref{eq:Hp}) gives us the evolution 
\begin{align*}
    e^{-i\left(f^x_o\sigma_o^x-f^z_o\sigma_o^z+\sum_iJ_i\sigma_i^z\sigma_o^z\right)}\ket{\bs{s}}\ket{0}_o &= \ket{\bs{s}}e^{-i\left(f^x_o\sigma_o^x-f^z_o\sigma_o^z+\sum_iJ_is_i\sigma_o^z\right)}\ket{0}_o\\
    =&\ket{\bs{s}}e^{-i\left(f^x_o\sigma_o^x+h(\bs{s})\sigma_o^z\right)}\ket{0}_o
\end{align*}
The evolution of the output is that of a rotation $e^{-i\omega \hat{\bs{n}}\cdot \bs{\sigma}}$ with $\hat{\bs{n}} = (f^x_o, 0, h(\bs{s}))/\omega$, $\omega= \sqrt{(f^x_o)^2+h(\bs{s})^2}$, and $\bs{\sigma} = (\sigma_o^x,\sigma_o^y,\sigma_o^z)$. This unitary can be represented as 
\begin{align*}
    \mathds{1}\cos(\omega)-i\hat{\bs{n}}\cdot\bs{\sigma}\sin\omega.
\end{align*}
Multiplying this last equation with the vector $\ket{0}_o$ gives us 
\begin{align}
    \ket{\bs{s}}\left( \left[\cos{\omega(\bs{s})}-i\frac{h(\bs{s})}{\omega(\bs{s})}\sin{\omega(\bs{s})}\right]\ket{0}_o\notag+i\frac{\Omega_o}{\omega(\bs{s})}\sin{\omega(\bs{s})}\right)\ket{1}_o.
\end{align}
If instead of a single ket $\ket{\bs{s}}$, we were given that the perceptron starts in the state $\sum_{\bs{s}}\psi(\bs{s})|\bs{s}\rangle|0\rangle_o$, since the unitary evolution is linear, we would obtain Eq.~(\ref{eq:PEvolved}). Note that according to the above,
\begin{equation}\label{eq:NativeF}
    f_{\text{QML}}(h(\bs{s})) = \arcsin{\frac{1}{\sqrt{1+h(\bs{s})^2/(f^x_o)^2}}\sin\left(f^x_o\sqrt{1+h(\bs{s})^2/(f^x_o)^2}\right)}
\end{equation}
which is an oscillatory function on $h(\bs{s})$ and $f^x_o$ and thus nonlinear on the input.\par

Now, we derive the activation function obtained from rotating the output qubit. Consider an activation function with a 
Fourier series valid on an interval $h\in [-\pi, \pi]$ 
\begin{align*}
    \fml(h) &= a_0 +\sum_{k=1}^\infty a_k \cos\left(kh\right)+\sum_{k=1}^\infty b_k \sin\left(kh\right), \\
    a_k  &= \frac{1}{\pi}\int_{-\pi}^{\pi}dh \fml(h)\cos({kh}), \\
    b_k  &= \frac{1}{\pi}\int_{-\pi}^{\pi}dh \fml(h)\sin({kh}).
\end{align*}
Accordingly, we can rewrite $\fml(h)\sigma_N^x$ as 
\begin{align*}
    \fml(h)&=a_0\sigma^x_N+\sum_k(a_k\cos kh +b_k\sin kh)\sigma_N^x \notag\\
    \fml(h)&=a_0\sigma^x_N+\sum_k\frac{a_k}{2}(\cos kh \sigma_N^x+\sin kh \sigma_N^y +\cos kh \sigma_N^x-\sin kh \sigma_N^y ) \notag\\
    &+\sum_k\frac{b_k}{2}(\sin kh \sigma_N^x+\cos kh \sigma_N^y +\sin kh \sigma_N^x-\cos kh \sigma_N^y )\notag\\
    &= a_0\sigma^x_N+\sum_k\frac{a_k}{2}(e^{-ikh/2\sigma_N^z}\sigma_N^xe^{ikh/2}+ e^{ikh/2\sigma_N^z}\sigma_N^xe^{-ikh/2}) \notag\\
    &-\sum_k\frac{b_k}{2}(e^{-i(kh/2+\pi/4)\sigma_N^z}\sigma_N^xe^{i(kh/2+\pi/4)}+ e^{i(kh/2+\pi/4)\sigma_N^z}\sigma_N^xe^{-i(kh/2+\pi/4)})
\end{align*}
Thus, one can write 
\begin{align*}
    U_{\fml} = e^{-ia_0\sigma_o^x}&\Pi_k \exp\left\{-i\frac{a_k}{2}\left(e^{-ikh/2\sigma_o^z}\sigma_o^x e^{ikh/2\sigma_o^z}+e^{ikh/2\sigma_o^z}\sigma_o^x e^{-ikh/2\sigma_o^z}\right)\right\}\notag\\
    &\Pi_k\exp\left\{i\frac{b_k}{2}\left(e^{-i(kh/2+\pi/4)\sigma_o^z}\sigma_o^x e^{i(kh/2+\pi/4)\sigma_o^z}+e^{i(kh/2+\pi/4)\sigma_o^z}\sigma_o^x e^{-i(kh/2+\pi/4)\sigma_o^z}\right)\right\}
\end{align*}
Let us denote $R_z(x)=e^{-ix/2\sigma^z_o}$, $H_{o} = \sigma_o^x$, and let $W_{k}(h), V_k(h)$ denote the exponentials above for $a_k$ and $b_k$ respectively;
\begin{align}
    W_k(h) &= \exp\left\{-i\frac{a_k}{2}\left(e^{-ikh/2\sigma_o^z}\sigma_o^x e^{ikh/2\sigma_o^z}+e^{ikh/2\sigma_o^z}\sigma_o^x e^{-ikh/2\sigma_o^z}\right)\right\}\\
    V_k(h) &= \exp\left\{i\frac{b_k}{2}\left(e^{-i(kh/2+\pi/4)\sigma_o^z}\sigma_o^x e^{i(kh/2+\pi/4)\sigma_o^z}+e^{i(kh/2+\pi/4)\sigma_o^z}\sigma_o^x e^{-i(kh/2+\pi/4)\sigma_o^z}\right)\right\} \\
    U_{\fml} &= e^{-ia_0\sigma^x_o}\Pi_k U_k(h)V_k(h). \label{eq:Uf_pulsed_appendix}
\end{align}
Note that the Hamiltonians above are of the form $H = e^{ih\sigma_o^z}\sigma^x_o e^{-ih\sigma_o^z}$. If we had white noise on $h\rightarrow h+\chi$ where $\chi\sim N(0,s_z)$, the average Hamiltonian is (to second order in $s_z$) 
\begin{equation}
    \bar{H} \approx (1-s_z^2)e^{ih\sigma^z_o}\sigma^x_oe^{-ih\sigma^z_o}.
\end{equation}
Thus, the Hamiltonian is robust to this error. \par
\begin{figure}
    \centering
    \includegraphics[width=0.8\linewidth]{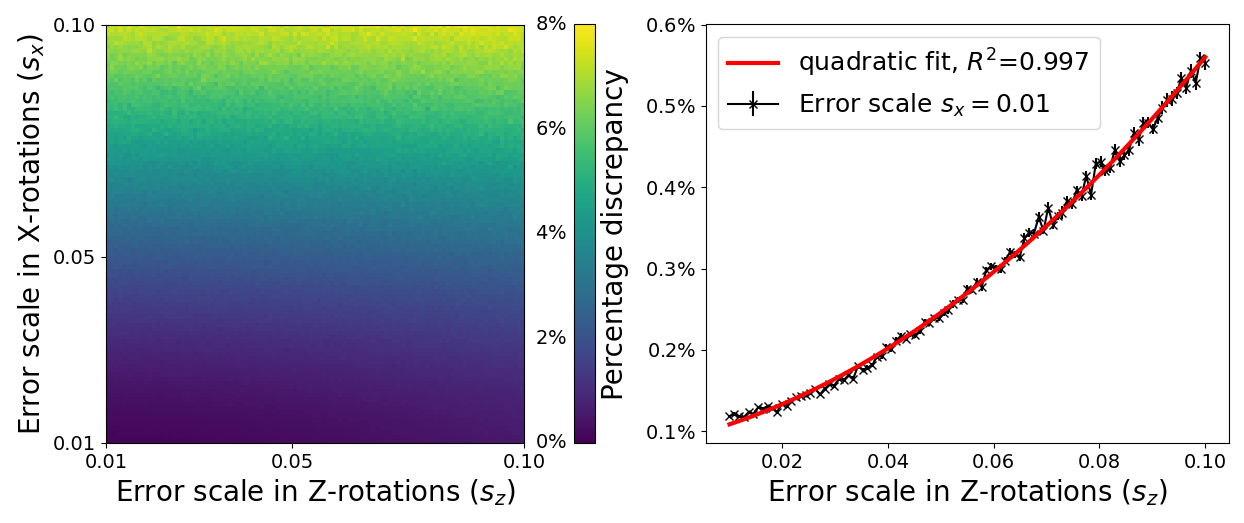}
    \caption{Percentage discrepancy of approximating the sigmoid function with 15 Fourier terms. The left panel shows the discrepancy as a function of error scale in each type of rotation. The discrepancy is largely independent to the error scale $s_z$. The right panel shows the percentage discrepancy for fixed $s_x=0.01$ and as a function of $s_z$. The error follows a quadratic prediction.}
    \label{fig:errors}
\end{figure}

 
Let us now consider introducing zero-mean Gaussian noise to the Fourier coefficients $a_k, b_k$ sampled from $N(0,s_x)$. We call $s_x$ and $s_z$ the error scales. Note that these labels highlight that the Fourier coefficients are related to evolving under $\sigma^x_o$ (x-rotations), while $h$ is related to evolving under $\sigma^z_o$ (z-rotations). Fig.~\ref{fig:errors} shows the percentage discrepancy of calculating the sigmoid function as a function of the error scales. The percentage discrepancy is calculated as the average of the quantity
\begin{equation}
    100\int_{-3}^3 dh\frac{|P_0(h)-\text{sigmoid}(h)|^2}{\text{sigmoid}(h)}.
\end{equation}
The average is taken over 50 instances of the errors on the coefficients and the field $h$. Note that here we chose the interval $h\in [-3,3]$ and not $h\in [-\pi,\pi]$. This choice is based on the observation that our approximation will largely fail near the endpoints of the interval $[-\pi,\pi]$. This discrepancy is due to the fact that any finite Fourier series fails most largely at the endpoints (which in our case is $\pm\pi$).

In Fig.~\ref{fig:errors} the left panel shows that the discrepancy is largely independent to the scale $s_z$ as expected. The right panel shows that the discrepancy follows a quadratic fit as a function of $s_z$ and for fixed $s_x$. 

\section{Quantum perceptrons and a universal gate set}\label{A:Universality}
\subsection{Supplementary calculation for identity gates}\label{A:UniIdentity}
In this Appendix, we provide a mathematical proof that a restricted version of the QP's Hamiltonian in Eq.~(\ref{eq:Hp}) can realize a universal gate set. For concreteness, we show that a QP contains the Clifford+T universal gate set~\cite{williams2011quantum}. We need to show that (i) a QP can realize the identity gate; (ii) a QP can entangle any two qubits $i,j$ through a control-Z gate where $CZ_{ij} = e^{i\pi\ket{1}\bra{1}_i\otimes\ket{1}\bra{1}_j}$, and that (iii) the QP can realize any single qubit rotations including the Hadamard- ($H$), phase- ($P$), and T-gates. \par

\begin{proof}[\textbf{Identity gate}] We here show that for any input qubit $i$ Eq.~(\ref{eq:Hp}) we can realize the trivial unitary $\mathds{1}_i$, where $\mathds{1}_i = \ket{0}\bra{0}+\ket{1}\bra{1}$ is the identity operator. To achieve this, we must eliminate the interactions between the input $i$ and the output. We do this using the time-dependent fields on qubit $i$. \par
To make this clear, let us consider a system that undergoes $N_r$ rotations (i.e., chirped field so that the fields dominate the dynamics over interaction terms) $\{R_n\}_{n=1}^{N_r}$. The leading-order average Hamiltonian $H_{av}$ is a weighted average 
\begin{equation}
    H_{av} = \sum_{k=1}^{N_r}\tau_k\tilde{H}_k,
\end{equation}
where $\tau_k$ is the duration between the $(k-1)^{th}$ and $k^{th}$ rotations, which are denoted $R_{k-1}$ and $R_k$. Here, $\tilde{H}_k=(R_{k-1} \hdots R_{1})^\dagger H_{\text{nat}}(R_{k-1}\hdots R_{1})$ is the toggling frame Hamiltonian. This paradigm of the toggling frame and native Hamiltonian is called average Hamiltonian theory~\cite{PhysRevX.10.031002}.\par 
We now apply this framework to achieve an identity gate. If we want input $i$ to decouple from the computation, we can rotate qubit $i$ by a series of $\pi$-rotations along the $x$-axis. That is, we choose $R_{k}=e^{-i\pi/2\sigma_i^x}$, and $\tau=1/N_r$. To see this, note that the native Hamiltonian between qubits $i$ and $o$ is $H_{s}=J_i\sigma^z_i\sigma^z_o$. But $R^\dagger_k\sigma_i^z R_k=-\sigma^z_i$ such that $R_k^\dagger H_{s} R_k = -H_{s}$. Rotating at each time increment $\tau$ produces the average Hamiltonian 
\begin{equation}\label{eq:DoNothing}
    H_{av} = \frac{1}{N_r}\left(\sum_{k \text{ even}} H_{s}+\sum_{k\text{ odd}}-H_{s}\right) = 0
\end{equation}
for even $N_{r}$. Thus, through rotation, we can realize identity gates on average.
We can also show that this rotation sequence produces an effective Hamiltonian with vanishing higher moments. To see this, we rely on average Hamiltonian theory (for a comprehensive review of AHT see Appendix A in~\cite{PhysRevX.10.031002}). According to AHT, the evolution of a time-dependent Hamiltonian $H(t)$ generates can be also generated by a time-independent Hamiltonian $U = e^{-iH_{\text{eff}}}$ where $H_{\text{eff}}\approx \sum_{k=1}^l H_l$ where $l$ is the truncation order and $H_k$ is the $k$th moment contribution. Given a rotation sequence, the average Hamiltonian, or fist moment, is
\begin{equation}
    H_{avg} = H_1 = \int_{0}^1 H(t)dt
\end{equation}
where $H(t)$ is the Hamiltonian at time $t$ generated by the rotation sequence. In the case of $R_{k}=e^{-i\pi/2\sigma_i^x}$, as shown in Eq.~(\ref{eq:DoNothing}), this average Hamiltonian vanishes. The next moment of this Hamiltonian is 
\begin{equation}
    H_2 = -\frac{i}{2}\int_{0}^1dt_2\int_0^1dt_1[H(t_2), H(t_1)]
\end{equation}
For the case of $R_{k}=e^{-i\pi/2\sigma_i^x}$ the commutators contain terms of the form $\pm [\sigma^z_i\sigma^z_o, \sigma^z_j\sigma^z_o]$ all of which vanish. Thus, $H_2=0$. Similarly, higher order moments contained nested commutators of Pauli-Z operators, all of which vanish. \par 
\end{proof}

\begin{proof}[\textbf{Two-qubit entangling gates}]
We now demonstrate that a QP can implement two-qubit entangling gates between any pair of qubits. Evidently, Eq.~(\ref{eq:Hp}) tells us that any input and the output can interact via an entangling gate. Suppose for a moment that only input $i$ interacts with the output. Introducing a rotation $R_z(\frac{\pi}{2}) = e^{-i\frac{\pi}{4}\sigma_i^z}$ and choosing $f^z_o=-\pi/4$, $J_i = -\pi/2$ with $f^x_o=0$ results in the evolution 
\begin{align}
    \exp\left\{-i\frac{\pi}{4}\left(\sigma_i^z+\sigma_o^z-\sigma_i^z\sigma_o^z\right)\right\} &= e^{-i\pi/4}e^{i\pi\ket{1}\bra{1}_i\otimes\ket{1}\bra{1}_o} \notag\\
    &=e^{-i\pi/4}CZ_{io}
\end{align}
where we have used the observation $2\ket{1}\bra{1}_i = \mathds{1}_i-\sigma_i^z$~\cite{gao2017quantum}. Thus, up to an arbitrary phase, a QP plus rotations entangles the input and the output via a CZ gate.\par
The inputs, however, do not directly interact via Eq.~(\ref{eq:Hp}). Since inputs $i,j$ interact with the output, they can interact with each other via a second-order interaction. \par

We now turn to the derivation of two-qubit gates. Suppose we have $f^z_o=0$ and 
\begin{equation}\label{eq:HPPerturbative}
    H_p = -f^x_o \sigma_o^x+\sum_{j}J_{j}\sigma^z_j\sigma^z_o= H_0+V.
\end{equation}
where $f^x_o\gg J_{j}$ so that $V$ is a perturbation. Let's construct the effective Hamiltonian on the input layer. We use the Schrieffer-Wolf transformation~\cite{bravyi2011schrieffer}. That is, we seek a unitary transformation $S$ so that $H'=e^{S}He^{-S}$ is diagonal to lowest order in $V$. We expect that since $V$ is small, in the sense that $|\lambda/f^x_o|\ll 1$ for any eigenvalue $\lambda$ of $V$, $S$ is also small. Using the Baker-Campbell-Hausdorff formula 
\begin{equation}
    H'=H_0+V+[S,H_0]+[S,V]+\frac{1}{2}[S,[S,H_0]]+O(V^3).
\end{equation}
If we choose $[S,H_0]=-V$, then 
\begin{equation}
    H'=H_0+\frac{1}{2}[S,V]+O(V^3)
\end{equation}
and thus to second-order perturbation theory, $[S,V]$ gives us the effective Hamiltonian. Suppose the eigenvectors and eigenvalues of $H_0$ are known so $H_0=\sum_{m}E_m|m\rangle\langle m|$. Then, $[S,H_0]=-V$ gives us 
\begin{align}
    \sum_m\left(SE_m|m\rangle\langle m|-E_m|m\rangle\langle m|S\right) &= -V\notag\\
    \sum_mE_{m}\left(\langle i|S|m\rangle\langle m|j\rangle-\langle i|m\rangle\langle m|S|j\rangle\right)&=-\langle i |V|j\rangle \notag\\
    \langle i|S|j\rangle &= \frac{\langle i|V|j\rangle}{E_i-E_j}.
\end{align}
This immediately gives us 
\begin{equation}
\frac{1}{2}[S,V]_{ij}=\frac{1}{2}\sum_k\left(\frac{\langle i|V|k\rangle\langle k|V|j\rangle}{E_i-E_k}-\frac{\langle i|V|k\rangle\langle k|V|j\rangle}{E_k-E_j}\right)
\end{equation}
which is the result of second-order perturbation theory: \begin{equation}\label{SWHami}
H'=H_0+\frac{1}{2}\sum_{ij}\sum_k\left[\frac{\langle i|V|k\rangle\langle k|V|j\rangle}{E_i-E_k}-\frac{\langle i|V|k\rangle\langle k|V|j\rangle}{E_k-E_j}\right]|i\rangle\langle j|  
\end{equation}
It is usual to ask that $V$ is completely off-diagonal in the eigenbasis of $H_0$. Notice that the eigenvectors of $H_0$ give us the basis $\{|+\rangle_o|\sigma\rangle, |-\rangle_o|\sigma\rangle\}_{\sigma}$ where $\{|\sigma\rangle\}$ is any eigenbasis on the input qubits. The energies are $-\Delta^x_o,\Delta^x_o$.
To see that the Hamiltonian in eq. Eq.~(\ref{SWHami}) is indeed diagonal, consider the case that $|i\rangle=|+_o,\sigma\rangle$ and $|j\rangle=|-_o,\sigma'\rangle$. The terms in the brackets in Eq.~(\ref{SWHami}), for $|k\rangle=|k_1,\sigma''\rangle$ with $k_1=\pm$, are easily seen to be zero since $\langle -_o|\sigma^z_o|-_o\rangle=\langle +_o|\sigma_o^z|+_o\rangle=0$. Thus, we end up with 
\begin{align}\label{EffectiveH}
    H'&=H_0-|+\rangle\langle+|_o\otimes\sum_{ij}\frac{J_{i}J_{j}}{f^x_o}\sigma^z_i\sigma^z_j+ |-\rangle\langle-|_o\otimes\sum_{ij}\frac{J_{i}J_{j}}{f^x_o}\sigma_i^z\sigma_j^z \notag\\
    &=|+\rangle\langle+|_o\otimes \left(-f^x_o+\sum_{ij}\frac{J_{i}J_{j}}{f^x_o}\sigma^z_i\sigma^z_j\right)+|-\rangle\langle -|_o\otimes \left(f^x_o-\sum_{ij}\frac{J_{i}J_{j}}{f^x_o}\sigma^z_i\sigma^z_j\right)\notag \\
    &=|+\rangle\langle+|_o\otimes H^{\text{in}}_{\text{eff}}-|-\rangle\langle-|_o\otimes H^{\text{in}}_{\text{eff}}.
\end{align}
This last equation leads to the result in Eq.~(\ref{eq:TwoQubitHamiltonian}). \par 
Lastly, consider the case where only qubits $i=1,2$ participate in this effective Hamiltonian and that after this Hamiltonian acts, we also rotate the inputs by an $R_z\left(\frac{-\pi}{2f^x_o}\right)$. Choosing $J_1 = -J_2 = \sqrt{\pi/4}$ we obtain the unitary 
\begin{align*}
    U_{12} &= R^1_z\left(\frac{-\pi}{2f^x_o}\right)R^2_z\left(\frac{-\pi}{2f^x_o}\right)e^{-if^x_o}e^{-i\frac{\pi}{4f^x_o}\sigma_1^z\sigma_2^z}\\
    &=e^{-if^x_o}e^{-i\frac{\pi}{4f^x_o}\left(\sigma_1^z+\sigma_2^z-\sigma_1^z\sigma_2^z\right)}\\
    &=e^{if^x_o}e^{-i\frac{-\pi}{4}CZ_{io}^{1/f^x_o}}.
\end{align*}
If we perform this operation $f^x_o$ times, then our gate is approximately $CZ_{io}$ up to the phase $\pi f^x_o/2-(f^x_o)^2$. Since $H_{\text{eff}}$ is correct up to order $\mathcal{O}((f^x_o)^{-2})$, the total error is of order $\mathcal{O}((f^x_o)^{-1})$.
\end{proof}
\begin{proof}[\textbf{Single-qubit gates}]
Finally, the construction of single-qubit gates follows from the presence of single-qubit time-dependent fields. In fact, to complete the Clifford+T gates, it suffices to use rotations of the form $R_{z}(\pi/4)$ and $R_x(\pi/4)$~\cite{kliuchnikov2015practical}.\par 
The Clifford gates are the single-qubit Hadamard- ($H$), phase- ($P$) gates, and the two-qubit control-Z ($CZ$) gate. We augment the set $\{H,P, CZ\}$ with the so-called $T$ gate to form a universal gate set~\cite{williams2011quantum}. The single-qubit gates are defined as follows
\begin{equation}
    H = \frac{1}{\sqrt{2}}\begin{pmatrix}
    1 & 1 \\
    1 & -1\end{pmatrix},\quad P = \begin{pmatrix}
    1 & 0 \\
    0 & i\end{pmatrix}, \quad T = \begin{pmatrix}
    1 & 0 \\
    0 & e^{i\pi/4}\end{pmatrix}.
\end{equation}
We use the convention $R_{x,z}(\theta) = e^{-i\theta/2\sigma^{x,z}}$. Evidently $P=T^2$. Note that $T = e^{-i\pi/8}R_z(\pi/4)$. Lastly, a straightforward calculation shows that $H = -iR_x(\pi/2)R_z(\pi/2)R_x(\pi/2)$. Since $R_{x,z}(2\theta)=R_{x,y}(\theta)^2$, rotations $R_{x,z}(\pi/4)$ are all we need to produce the Clifford+T set. As long as those single-qubit time-dependent fields are present in a QP, the QP can realize any quantum computation.
\end{proof}

\section{Details on the numerical implementations of QPs}\label{A:numerics}

To simulate the Hamiltonian in Eq.~(\ref{eq:Hp}), we use the method of troterization: 
\begin{equation}
    U(T;\bs{\theta})\approx \prod_{k=1}^{P}e^{-i\tau \sum_{n,\alpha}f_n^{\alpha}(t_k,\bs{\Delta})\sigma_n^\alpha}e^{-i\tau H_{\text{nat}}(\bs{J})} = \prod_{k=1}^{P} U_{\text{ctr}}(t_k;\bs{\Delta})U_{\text{nat}}
\end{equation}
where $\tau=T/P$. This approximation is valid up to second order in $\tau$. Moreover, each unitary above $U_{\text{nat}}$ and $U_{\text{ctr}}$ have a matrix product state (MPS) representation. ~\cite{vidal2004efficient, paeckel2019time}. An MPS represents a quantum state as the contraction of tensors 
\begin{equation}\label{eq:MPS}
    |\psi\rangle= \sum_{\bs{s}}\left(A^{s_1}_{i_1}A^{s_2}_{i_1i_2}\hdots A^{s_{N}}_{i_{N-1}}\right)|s_1s_2\hdots s_N\rangle
\end{equation}
where $\bs{s}$ is a binary vector of length $N$, $A^{s_1}_{i_1}$ and $A^{s_N}_{i_N}$ are vectors of size $d\times 1$ and $1\times d$ respectively, and matrices $A^{s_k}_{i_k}$ for $1<k<N$ are matrices of size $d\times d$.The parameter $d$ is called bond-dimension, and it physically corresponds to the degree of entanglement shared between qubits $k$ and $k\pm 1$. It is a convention to choose $D$ to be uniform for all matrices. We can denote $A^{[k]}$ as the 3-tensor of dimension $2\times d\times d$ containing information about qubit $k$. Note that these tensors have a dimension of size 2 which corresponds to the physical index (or leg) of the qubit, and two dimensions of size $d$ which correspond to virtual legs to neighboring qubits. These virtual legs contain information about the entanglement structure of the state. While a state $|\psi\rangle$ would normally be described by the coefficients $\psi(\bs{s})$ for each basis state, the matrices in Eq.~(\ref{eq:MPS}) can be found from these coefficients via Smidth-decomposition~\cite{schollwock2011density}.  \par
Using similar logic, operators can also be written in terms of matrix products leading to the notion of a matrix product operator (MPO)
\begin{equation}
    \hat{\mathcal{O}} = \sum_{\bs{s}, \bs{c}}M_{i_1}^{s_1,c_1}M_{i_1,i_2}^{s_2,c_2}\hdots M_{i_{N-1}}^{s_{N},c_N}|s_1s_2\hdots s_N\rangle\langle c_1c_2\hdots c_N|.
\end{equation}
The tensors $M^{[k]}$ are now $2\times2\times d_{\text{op}}\times d_{\text{op}}$ dimensional (with a bond dimension $d_o$ possibly different than that of an MPS) containing two physical indexes and two virtual indexes. The virtual indexes contain information as to how the operator entangles neighboring qubits. MPOs can be applied to an MPS producing a new MPS with new bond-dimension $dd_{\text{op}}$. The process of propagating an MPS through a series of $P$ operators takes run-time $\mathcal{O}(Nd(d_{\text{op}})^P)$~\cite{vidal2004efficient}. \par 
Let us now apply the MPS framework to our QPs. First of all, single-qubit operations in $U_{\text{ctr}}(t_k)$ can be mapped into MPOs with $d_{\text{op}}=1$. This follows because $U_{\text{ctr}}(t_k)$ is a product over single qubit rotations:
\begin{equation*}
    U_{\text{ctr};\bs{\Delta}}(t_k) = \prod_{n=1}^{N+1}e^{-i\bs{h}_n(t_k;\bs{\Delta})\cdot \bs{\sigma}_n} = \prod_{n}R_{n}(t_k).
\end{equation*}
For a rotation, $M^{s_k,c_k} = \langle c_n| R_n |s_n\rangle$ is the matrix element of the rotation (a scalar). \par 
Secondly, and more interestingly, evolving under $H_{\text{nat}}$ can also be done via first mapping the Hamiltonian into an MPO. Here we must remark that not all Hamiltonians have exact MPO representations. Moreover, $H_{\text{nat}}$ is, in some sense, a long-range Hamiltonian since qubit $i=1$ interacts with the output which in the MPS language is $N$ sites away. However, this apparent long-range is due to the linear geometry of the MPS form, and physically all input qubits are neighbors to the output. This realization is what allows us to derive the exact MPO form
\begin{equation}
    H_{\text{nat}}^{[1]} = \begin{pmatrix}
    \mathds{1}_1 & J_1\sigma_1^z
    \end{pmatrix}, \quad H_{\text{nat}}^{[k]} = \begin{pmatrix}
    \mathds{1}_k & J_k\sigma_k^z\\
    0 & \mathds{1}_k
    \end{pmatrix}, \quad  H_{\text{nat}}^{[o]} = \begin{pmatrix}
    0 \\
    \sigma_o^z\\
    \end{pmatrix}.
\end{equation}
Note that with this notation $H_{\text{nat}} = H^{[1]}H^{[2]}\hdots H^{[o]}$. The bond dimension of $H_P$ is $d_{\text{op}}=2$. It has been shown that for a Hamiltonian with an MPO, its corresponding unitary evolution, $U = e^{-i\tau H_{\text{nat}}}$ in our case, can be approximated by an MPO as well~\cite{paeckel2019time}. In particular, we use the approximation valid to second order in $\tau$:
\begin{equation}
    U_{\text{nat}}\approx 1-i\tau H_{\text{nat}}-\frac{\tau^2}{2}H^2_{\text{nat}}.
\end{equation}
The multiplication of two MPOs of bond dimension $d_{\text{op}}$ results in an MPO of bond dimension $d_{\text{op}}^2$, while addition results in adding the bond dimensions. So our approximation has a bond dimension $1+d_{\text{op}}+d_{\text{op}}^2\sim \mathcal{O}(d_{\text{op}}^2)$. Thus, the algorithm scales as $\mathcal{O}(Nd_{\text{op}}^{2P})$. The error in the MPO approximation scales as $\mathcal{O}(P\tau^3)$, and the error in the trotterization as $\mathcal{O}(P\tau^2)$.\par 

In Sec.~\ref{Sec:energy}, we consider the effective Hamiltonian in Eq.~(\ref{eq:EffectiveTrotter}).
Since we know that Heisenberg Hamiltonians, $H_{\text{Heis}}\propto \sum_{ij,\alpha}\sigma^\alpha_i\sigma^\alpha_j$, tend to have the largest energy separation between ground state and separable states, we give our QP the number of pulses needed to achieve the Heisenberg Hamiltonian as the effective Hamiltonian with pulses. We could imagine using $P=3n+1$ pulses where the pulses take $Z\rightarrow X\rightarrow Y\rightarrow Z$ and so on. Then, the average effective Hamiltonian is $\frac{J^2}{3\Omega_o}\sum_{ij,\alpha}\sigma^\alpha_i\sigma^\alpha_j.$ It is worth noting that higher orders of the effective Hamiltonian are also Heisenberg-type since they are composed of nested commutators all of which yield Heisenberg operators. For example, the second order term is composed of terms coming out of commutators like
\begin{equation*}
[\sigma^\alpha_i\sigma^\alpha_j,\sigma_k^\beta\sigma_l^\beta] = (\delta_{ik}\delta_{jl}+\delta_{il}\delta_{jk})(2i\epsilon_{\alpha\beta\gamma})^2\sigma_i^\gamma \sigma_j^\gamma.
\end{equation*}

The approximation of the average Hamiltonian to a Heisenberg Hamiltonian increases with an increasing number of pulses. Thus, one could increase $n$ while balancing that $(J/(f^{x}_o)^2)(3n\tau+1)\ll 1$. This means that $P=3n+1$ is a good heuristic for how many time-steps are needed. In the case of our application, we used $n=1$.

\section{Barren plateaus and entanglement dropout in quantum perceptrons}\label{A:ED}
\begin{figure}[b]
\includegraphics[width=0.8\linewidth]{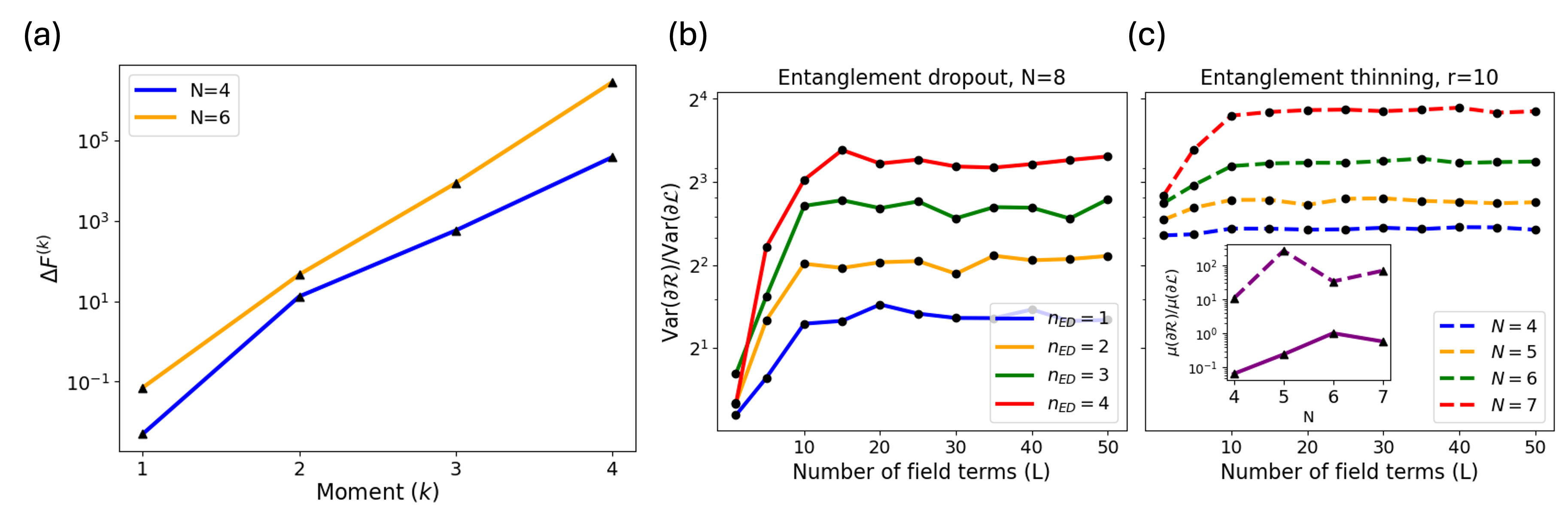}
\caption{(a) Frame potentials for the QPs under at $L=50$ for different input qubit numbers. (b) Improvement of the gradient's variance using entanglement dropout. (c)Improvement of the gradient's variance using entanglement thinning. Inset shows that the average gradient remains unchanged with entanglement dropout while it improves with entanglement thinning}
\label{fig:FramePotetentials}
\end{figure}

A variational ansatz with parameters $\bs{\theta}$ initialized at various random parameters $M=\{\bs{\theta}_i^{(0)}\}_i$ forms a unitary ensemble $\{U(\bs{\theta}_i^{(0)})\}_i.$ A barren plateau (BP) is defined as the zeroing out of the gradient in average $M$ together with an exponentially small gradient variance.\par 
BPs necessarily appear when the unitary ensemble is a $k$-design for $k\geq 2$ (i.e., $M$ has the same fist $k$ moments as a Haar random unitary ensemble). Avoiding BPs necessitates a deviation of a variational ansatz's statistics from those of a $k$-design, but such deviation alone may not completely remove the barren plateaus. To measure such deviation, the frame potential has been proposed as a measure of how close a unitary ensemble is to a $k$-design under the diamond norm~\cite{roberts2017chaos}. For an ensemble of unitaries $M$ of size $|M|$ the frame potential is defined by
\begin{equation}
    F^{(k)}(M) = \frac{1}{|M|^2}\sum_{U,V\in M}\left|\text{Tr}\left(U^\dagger V\right)\right|^{2k},
\end{equation}
and for a $k$-design $F^{(k)}=k!$. Fig.~\ref{fig:FramePotetentials} shows the difference in the frame potentials between the ensemble of our randomly initialized QPs and those of a $k$-design (i.e., $\Delta F^{(k)}\equiv F^{(k)}(\text{QPs})-k!$). Large values of $\Delta F^{(k)}$ indicate that QPs fail to resemble $k$-designs for all $k\geq 2$. However, as shown in Fig.~\ref{fig:Trainability}(a), QPs do exhibit BPs. This again shows that $k$-design resemblance is sufficient but not necessary for a VQA to exhibit barren plateaus. We thus argue in the main text that barren plateaus arise due to entanglement build-up in the QP's star-like architecture.\par 
One way to mitigate barren plateaus is entanglement dropout (ED). To do this ``dropping out" of qubits, we follow a recently introduced technique called
\textit{entanglement dropout} (ED)~\cite{kobayashi2022overfitting} which resembles dropout techniques introduced in classical machine learning~\cite{srivastava2014dropout}. For ED, we define a probability function $\boldsymbol{p}$ depending on a set of parameters $\boldsymbol{\alpha}$ which determines the probability of sampling binary vectors $\boldsymbol{z}$ such as $(1,1,...), (1,0,...),$ etc. For a given sampling of $\boldsymbol{z}$ we let the perceptron evolve with weights $\boldsymbol{J}\cdot\boldsymbol{z}$ which drops out those qubits for which the entry in the dropout configuration, $\boldsymbol{z}$, is zero. This amounts to a regularized loss function given by
\begin{equation}
    \mathcal{R}_{\text{ED}}(\boldsymbol{\alpha}, \boldsymbol{J}) = \sum_{\boldsymbol{z}}\boldsymbol{p}(\boldsymbol{z}|\boldsymbol{\alpha})\mathcal{L}(\boldsymbol{J}\cdot\boldsymbol{z}).
\end{equation}
Thus, to perform gradient descent of this problem, one must compute the gradients $\partial_{\boldsymbol{J}}\mathcal{R}$ and $\partial_{\boldsymbol{\alpha}}\mathcal{R}$. In practice, this optimization requires computing the loss $\mathcal{L}(\boldsymbol{J}\cdot\boldsymbol{z})$ over an exponential number of dropout configurations $\boldsymbol{z}$. To avoid this exponential overhead, one can resort to doing Monte Carlo sampling of the loss. \par 
Fig.~\ref{fig:FramePotetentials}(b) shows the ratios of the variance of the loss' gradients with and without dropout for the case of $N=8$ for 500 random initialization of $\boldsymbol{J}$ and with different number of qubits $n_{\text{ED}}$ dropped out. Note that for the problem in~(\ref{eq:CHSH}), only the first input qubit should be active while the other qubits are irrelevant. Numerically, we observe that the variance of $\partial_{\boldsymbol{J}} \mathcal{R}_{\text{ED}}$ is exponentially larger than those without dropout. Below, we proof this result rigorously assuming that the active part of the QP exhibit a barren plateau. Therefore, we extend the results in~\cite{patti2021entanglement} to the case of entanglement dropout and the QP's architecture. Fig.~\ref{fig:FramePotetentials}(c) shows the variance improvement for entanglement thinning (ET) as described in the main text in Fig.~\ref{fig:Trainability}(b). The inset shows the mean value of the gradients for ED (solid line)and ET (dotted line). ED does not improve the mean values of the gradients while ET does. \par 
We now show that ED improves the variance of the gradients. Note that 
\begin{equation}
\partial_{\boldsymbol{J}}\mathcal{R}_{\text{ED}} = \sum_{M=1}^{N}\sum_{|\boldsymbol{z}|=M}p(\boldsymbol{z}|\boldsymbol{\alpha})\partial_{\boldsymbol{J}}\mathcal{L}(\boldsymbol{z}\cdot\boldsymbol{J}).
\end{equation}
When $|\boldsymbol{z}|=M$, the density matrix of the perceptron can be written as $\rho_A\otimes \rho_B$ where $A$ is the subsystem for the qubits where $z_i=1$. Moreover $\rho_B$ contains no entanglement. In Ref.~\cite{patti2021entanglement}, it was shown that for such factorized system, a barren plateau would obey the scaling law $2^{-2|A|}$ (See Eq.(18) in Ref.~\cite{patti2021entanglement}). Thus, even if we assume that the active subsystem during ED exhibits a barren plateau, we obtain 
\begin{align}
    \langle \partial_{\boldsymbol{J}}\mathcal{R}_{\text{ED}} \rangle_{\boldsymbol{J}} &= 0 \\
    \langle\left(\partial_{\boldsymbol{J}}\mathcal{R}_{\text{ED}}\right)^2\rangle_{\boldsymbol{J}} &\sim  \sum_{M=1}^{N}\sum_{|\boldsymbol{z}|=M}2^{-2|M|} p(\boldsymbol{z}|\boldsymbol{\alpha}).
\end{align}
For the sake of Fig.~\ref{fig:Trainability}(d), we picked $p$ to be nonzero only when $|\boldsymbol{z}|=n_{\text{ED}}$. In such case, $\text{Var}(\partial_{\boldsymbol{J}}\mathcal{R}_{\text{ED}})\sim 2^{-2n_\text{ED}}$. 

\section{Supplementary calculations on using QPs for quantum metrology}\label{A:metrology}
In this section, we explain how to measure the signal of Sec.~\ref{Sec:metrology} using the input qubits. We can also do it by measuring the output qubit alone. Suppose that after the circuit in Fig.~\ref{fig:metrology_circuit}, and before the measurement, the QP's state is 
\begin{equation}
    \sum_{\bs{y}}\psi(\bs{y})|\bs{y}\rangle |-\rangle
\end{equation}
where $|\bs{y}\rangle=|y_1,y_2,...,y_N\rangle$ is a separable state where $\sigma_i^y|y_i\rangle=y_i|y_i\rangle$. Now, suppose that we first rotate all the input qubits by $R_x(\pi/2)$ resulting in $\sigma_i^z\rightarrow \sigma_i^y$. Then using an $H_P$ layer with $f^x_o=f^z_o=0$ and $J_i=J$ we obtain 
\begin{equation}
    \sum_{\bs{y}}\psi(\bs{y})|\bs{y}\rangle |-\rangle\rightarrow \sum_{\bs{y}}\psi(\bs{y})|\bs{y}\rangle e^{-iJ\tau\sum_iy_i\sigma_o^z}|-\rangle_o
\end{equation}
Noticing that $\sum_i y_i = 2\langle\bs{y}| S_y|\bs{y}\rangle$ we see that the probability of observing the output neuron in the state $|+\rangle_o$ is
\begin{equation}
    P_{+}=\sum_{\bs{y}}\sin^2(2Jt\langle \bs{y}|S_y|\bs{y}\rangle)|\psi(\bs{y})|^2.
\end{equation}
For small $NJt\ll 1$ we obtain to second order in $t$
\begin{equation}
    P_{+} = (2Jt)^2\langle S_y \rangle^2 =(2Jt)^2(m\varphi N/2)^2.
\end{equation}
Therefore, with a circuit performing a global rotation on the input and a small-time evolution under $H_P$, we can use the output qubit to measure the signal $\varphi$ and the amplification $m$.

\end{document}